\documentclass[12pt,authoryear]{elsarticle}
 \usepackage{amssymb}
 \usepackage{natbib}
\usepackage{amsfonts,amsmath,epsfig,psfrag}
 \usepackage{amsmath}
 \usepackage{tikz}
 \usepackage{float}
\usepackage{pgflibraryshapes}
 \usepackage{graphicx,psfrag}
\usepackage[normalem]{ulem}
 \makeatletter \oddsidemargin 0.0cm \evensidemargin 0.0cm
\newcommand{\be}{\begin{equation}}
\newcommand{\en}{\end{equation}}
\newcommand{\la}{\label}
\newcommand{\ep}{{\varepsilon}}
\newcommand{\paa}{\partial}
\def\rr#1{(\ref{#1})}
\def\bm#1{\mbox{\boldmath{$#1$}}}

\def\ii{{\rm i}}

\newcommand{\s}[1]{{\Large\textsf{\textbf{#1}}}}

\begin{document}

\begin{frontmatter}
\title{\s{Axisymmetric necking versus Treloar-Kearsley instability in a hyperelastic sheet under equibiaxial stretching}}
\author[mymainaddress]{Mi Wang\corref{contrib}}
\author[secondaryaddress]{Lishuai Jin\corref{contrib}}
\author[thirdaddress]{Yibin Fu\corref{mycorrespondingauthor}}
\cortext[mycorrespondingauthor]{Corresponding author: y.fu@keele.ac.uk}
\cortext[contrib]{Authors contributed equally}
\address[mymainaddress]{Department of Mechanics, Tianjin University, Tianjin 300072, China}
\address[secondaryaddress]{Department of Materials Science and
Engineering, University of Pennsylvania, Philadelphia, PA 19104, USA}
\address[thirdaddress]{School of Computing and Mathematics, Keele University, Staffs ST5 5BG, UK}

\begin{abstract}
We consider bifurcations from the homogeneous solution of a circular or square hyperelastic sheet that is subjected to equibiaxial stretching under either force- or displacement-controlled edge conditions. We derive the condition for axisymmetric necking and show, for the class of strain-energy functions considered, that the critical stretch for necking is greater than the critical stretch for the Treloar-Kearsley (TK) instability and less than the critical stretch for the limiting-point instability. An amplitude equation for the bifurcated necking solution is derived through a weakly nonlinear analysis and is used to show that necking initiation is generally sub-critical. Abaqus simulations are conducted to verify the bifurcation conditions and the expectation that the TK instability should occur first under force control,  but when the edge displacement is controlled the TK instability is suppressed, and it is the necking instability that will be observed. It is also demonstrated that axisymmetric necking follows a growth/propagation process typical of all such localization problems.
\end{abstract}
\begin{keyword}
Necking \sep Treloar-Kearsley instability \sep equibiaxial tension\sep elastic localization\sep  nonlinear elasticity.

\vspace{0.5cm}

\noindent This paper is dedicated to the memory of Professor Hui-Hui Dai.
\end{keyword}
\end{frontmatter}

\section{Introduction}
Necking is a phenomenon commonly observed in the tension test of ductile metals. It has also been observed in soft materials such as semi-crystalline and glassy polymers \citep{ch1932, wa1967, cz1974, zc1974, ward2013, zb2018}, and hydro-gels \citep{ntk2006, mp2010, zhao2012}. Numerical simulations have been conducted by \citet{chen1971}, \citet{ne1972}, \citet{nm1978}, \citet{bn1979}, and \citet{si1988}, and there exists a large body of literature devoted to the analysis of necking based on 1D models; see, for instance, \citet{ba1964}, \citet{antman1972}, \citet{antman1973}, \citet{er1975}, \citet{ta1986}, \citet{owen1987}, \citet{coleman1988}, \citet{cn1988}, \citet{db2006}, \citet{dh2008}, \citet{dp2012}, \citet{ah2016}, and \citet{la2020}.

Under the framework of nonlinear elasticity, the initiation condition for necking was traditionally derived as the condition for a periodic mode satisfying the end conditions to appear, in the same manner as for a buckling mode under compression; see, for instance, \citet{wes1962}, \citet{hm1974}, \citet{hh1975}, \citet{ac1977},  \citet{st1998}, and \citet{tr2007}. However, there also exists an alternative point of view based on the dynamical systems theory whereby localisation is viewed to initiate from a \lq\lq zero wavenumber" mode \citep{ki1982, mi1991, fu2001}. For a simple and purely algebraic demonstration of how a localised mode bifurcates from a uniform state, we refer to \citet{fpl2008}. Although the critical loads associated with the two approaches are usually close to each other, the post-buckling behaviors predicted by the two approaches become qualitatively different. Whereas the \lq\lq zero wavenumber" point of view predicts a post-bifurcation profile that is more and more localised away from the bifurcation point, exactly as observed in numerical simulations and experiments, the other approach predicts a periodic profile that has a vanishingly small range of validity \citep{wf2021}. The \lq\lq zero wavenumber" approach has recently been applied to study elasto-capillary necking and has been shown to give predictions in excellent agreement with numerical simulations \citep{xuan2017, giudici2020, fjg2021}. This approach has previously been applied to study localised bulging in inflated rubber tubes; see
\citet{fl2016}, \citet{ylf2019}, \citet{wg2019}, and the references therein.

The main motivation for the current study is to provide insight into necking of a dielectric membrane that is subjected to the combined action of an electric field and mechanical forces. This phenomenon has been addressed in the papers by \citet{pz2012}, \citet{zd2017}, \citet{fdx2018}, and \citet{fxd2018}, but has otherwise received very little attention in the vast literature on electroelasticity. Instead, most literature on stability and bifurcation has focussed on wrinkling \citep{bg2011, do2014, yzs2017, sb2018, bs2018, do2019, xsc2021} or marginal violation of the Hessian criterion \citep{zs2007, no2008, lm2020, cy2021}.

To focus on the explanation of ideas without the extra complications arising from the incorporation of electric effect, we consider in this paper the purely mechanical case when a circular hyperelastic sheet is subjected to an all-round tension along its curved edge. The resulting trivial primary deformation is the same as in a square sheet that is subjected to equibiaxial tension. Such a square sheet is known to be susceptible to both Treloar-Kearsley (TK) and limiting-point instabilities. The former refers to the existence of deformation field in which the two in-plane stretches become unequal although the associated tensions are equal, whereas the latter refers to the fact that the tension as a function of the stretch on the trivial loading path may reach a maximum. The TK instability was first noted by \citet{ke1986} based on the experimental results of \citet{tr1949}, with its plane-strain counterpart analysed slightly earlier by \citet{Og1985}. It has subsequently been studied by \citet{mac1986}, \citet{og1987}, \citet{chen1987}, \citet{mc1992}, \citet{bm2005},  and \citet{st2007}.
However, our focus in the current study will be on a third instability, that associated with axisymmetric necking. Based on our previous studies on localised bulging of an inflated tube, it is tempting to conjecture that the bifurcation condition for necking might be that the Jacobian determinant of the two stretching forces, as functions of the two corresponding stretches, is zero on the trivial loading path (i.e. when the two in-plane stretches are equal). However, it turns out that the Jacobian determinant equal to zero may always be factorised into two separate equations, and each equation may have zero, one or two roots for the stretch depending on the material model used. When each equation has at least one root (which is the case for the class of material models considered), the two smallest roots correspond to the TK and limiting-point instabilities, respectively. Guided by our previous studies on the localised bulging of inflated rubber tubes, we postulate that localised necking may occur when a static axially symmetric extensional mode may exist in the limit of zero wavenumber (the extensional mode is the mode whose out-of-plane displacement is symmetric with respect to the mid-plane). The associated critical stretch then lies between the two smallest roots of the Jacobian determinant, and this is verified by Abaqus simulations.

The rest of this paper is organised as follows. After summarizing the governing equations in the next section, we present a linear analysis in Section 3 where we derive the bifurcation conditions for the static and axially symmetric, extensional and flexural modes, respectively. Each bifurcation condition is an implicit equation relating the stretch to the wavenumber in the radial direction. As for localised bulging in inflated rubber tubes, the initiation condition for necking is postulated to be the condition for the extensional mode with zero wavenumber to exist. This condition is further discussed in relation to the TK and limiting-point instabilities. In Section 4, we conduct a weakly nonlinear analysis and derive the amplitude equation in a small neighbourhood of the bifurcation point and show that the necking instability is generally sub-critical. In Section 5, we conduct Abaqus-based numerical simulations of the TK and necking instabilities and show that for the class of constitutive models considered, TK instability appears first under force control and it gives way to necking instability under displacement control. Section 6 is devoted to an analytical description of necking propagation. We determine the solutions for the two \lq\lq phases" analytically and compare them with numerical simulations. The paper is then concluded in the final section with some additional remarks.
%
%


\section{Governing equations}
We consider a sufficiently large circular or square hyperelastic sheet that is subjected to an equibiaxial tension in its plane.
We denote by $B_0$ and $B_e$ the undeformed and the finitely deformed configurations, respectively. The sheet thicknesses in $B_0$ and $B_e$ are denoted by $H$ and $h$, respectively. We shall look for axisymmetric deformations, and so cylindrical polar coordinates will be employed, with coordinates $\theta$, $z$ and $r$ in $B_e$ corresponding to 1-, 2-, and 3-directions, respectively. The $r$ and $\theta$ coordinates define positions in the plane, whereas $z$ measures distance in the out-of-plane direction such that $z=0$ corresponding to the mid-plane and the surfaces to $z=\pm h/2$. Thus, corresponding to the equibiaxial deformation, the three principal stretches are given by
\be \lambda_1=\lambda_3=\lambda, \;\;\;\; \lambda_2=\lambda^{-2} \la{las} \en
in terms of a single parameter $\lambda$ due to the constraint of incompressibility.

To introduce the incremental equations,  we denote the deformation gradient from $B_0$ to $B_e$ by $\bar{F}$ and the associated nominal stress by $\bar{S}$.  A small amplitude axially symmetric perturbation is now applied to $B_e$, giving rise to the final configuration $B_t$, and the associated incremental displacement field $\delta{\bm x}$ is given by
\be \delta{\bm x}=u(r, z) {\bm e}_r+v(r, z) {\bm e}_z, \la{incrr} \en
where ${\bm e}_r$ and ${\bm e}_z$  are the basis vectors in the $r$- and $z$-directions, respectively, and $u$ and $v$ are the associated displacement components.
The deformation gradient corresponding to the deformation  $B_0 \to B_t$ is denoted by $F$ and the associated nominal stress by ${S}$. We write $F=(I+\eta) \bar{F}$ so that $\eta$ denotes the deformation gradient associated with the incremental deformation $B_e \to B_t$. The divergence operator with respect to coordinates in $B_0$ and $B_e$ will be denoted by Div and div, respectively.

The incremental equilibrium equation can best be expressed in terms of the incremental stress tensor $\chi$ defined by
\be \chi^{\rm T}= \bar{J}^{-1} \bar{F} (S-\bar{S}), \la{incrs} \en
where the superscript T stands for transpose, and $\bar{J}$ denotes the determinant of $\bar{F}$ (which is unity in the current case but is kept in the formula to maintain the generality of the formula). With the use of the identity ${\rm div}\, \bar{J}^{-1} \bar{F}={\bm 0}$ and the equilibrium equations  ${\rm Div}\, \bar{S}={\bm 0}$ and ${\rm Div}\, {S} ={\bm 0}$, we obtain the incremental equilibrium equation
\be {\rm div}\, \chi^{\rm T}={\bm 0}. \la{incre} \en
For the current axisymmetric deformation, only the equations corresponding to $i=2, 3$ are not satisfied automatically, and they are given by
\be
\chi_{3j,j}+\frac{1}{r} (\chi_{33}-\chi_{11})=0, \;\;\;\; \chi_{2j,j}+\frac{1}{r} \chi_{23}=0. \la{incre1} \en
For our linear and weakly nonlinear analyses, we need expansions of $\chi_{ij}$ up to the quadratic order, and they are given by
\be
\chi_{ij}=B_{jilk} \eta_{kl}+\bar{p}  \xi_{ji} -p^* (\delta_{ji}-\xi_{ji})+\frac{1}{2} B_{jilknm}^2 \eta_{kl}\eta_{mn}+\cdots,
\la{incre2} \en
where
\be \xi_{ji}=\delta_{ji}-F^{-1}_{Ai} \bar{F}_{jA}= \eta_{ji}-\eta_{jm} \eta_{mi}+\cdots. \la{xxi} \en
See, e.g., \citet{fo1999}.
In \rr{incre2} the $\bar{p}$ and $p^*$ are the Lagrangian multipliers associated with the deformations $B_0 \to B_e$
and $B_e \to B_t$, respectively. The $\eta_{kl}$ are the components of $\eta$ given by
\be
 { \eta}=\left[\begin{array}{ccc} \dfrac{u}{r} & 0 &0 \\ 0 & v_z & v_r \\ 0 & u_z & u_r \end{array} \right], \;\;\;\; v_z \equiv \frac{\paa v}{\paa z}, \;\;\;\;v_r \equiv \frac{\paa v}{\paa r}\;\; {\rm etc}, \la{incre3} \en
 and $B_{jilk}$ and $B^2_{jilknm}$ are the 1st- and 2nd-order instantaneous elastic moduli, the expressions of which can be found in \citet{og1984} or \citet{fo1999}.


Due to the introduction of the extra variable $p^*$, the equilibrium equations are augmented by the incompressibility condition which can be expanded as
 \be \eta_{ii}-\frac{1}{2} \eta_{mn} \eta_{nm}+ ... =0. \la{incre4} \en
For the current problem, both the top and bottom surfaces are traction-free. Thus, the associated incremental boundary conditions  are
\be \chi_{22}=0,\;\;\;\;\chi_{32}=0,  \;\;\;\;{\rm on}\;\;  z= \pm h/2. \la{incre5a} \en


\section{Linear analysis}
With linearization, the incompressibility condition \rr{incre4} for the current problem reduces to
\be \frac{\paa\, (r u)}{\paa r}+\frac{ \paa\, (r v)}{\paa z}=0. \la{3.1} \en
It is then convenient to satisfy this condition automatically by introducing a \lq stream function\rq $\,\phi(r, z)$ such that
\be u=\frac{1}{r} \phi_z, \;\;\;\; v=-\frac{1}{r} \phi_r, \la{3.2} \en
where as in \rr{incre3} a subscript signifies differentiation (e.g. $\phi_z=\paa \phi/\paa z$).
The non-zero stress components are given by
\begin{eqnarray} \chi_{11}&=&(B_{1111}+\bar{p}) \frac{u}{r} +B_{1122} v_z+B_{1133} u_r-p^*, \noindent \\
  \chi_{22}&=& B_{1122} \frac{u}{r} +(B_{2222}+\bar{p}) v_z+B_{2233} u_r-p^*, \noindent \\
  \chi_{33}&=& B_{1133} \frac{u}{r} + B_{2233} v_z+(B_{3333}+\bar{p}) u_r-p^*, \noindent \\
  \chi_{23}&=& B_{3232} v_r+(B_{3223}+\bar{p}) u_z, \noindent \\
  \chi_{32}&=& B_{2323} u_z+(B_{2332}+\bar{p}) v_r. \noindent \end{eqnarray}
On substituting these expressions together with \rr{3.2} into the linearized forms of \rr{incre1} and then eliminating $p^*$ by cross-differentiation, we obtain
\be
\alpha \left( \phi_{rrrr}-\frac{2}{r} \phi_{rrr}+\frac{3}{r^2} \phi_{rr}-\frac{3}{r^3} \phi_{r}  \right)+2 \beta \left( \phi_{rrzz}-\frac{1}{r} \phi_{rzz} \right)+\gamma \phi_{zzzz}=0, \la{3.4} \en
where
\be \alpha=B_{1212}, \;\;\;\; 2 \beta=B_{1111}+B_{2222}-2 B_{1122}-2 B_{1221}, \;\;\;\; \gamma=B_{2121}. \la{3.5} \en
Note that these constants are the same as those in \cite{do1990} and \cite{fr1994} for plane-strain deformations. The $\alpha$, for instance, is given by
\be \alpha=\frac{(\sigma_1-\sigma_2) \lambda_1^2}{\lambda_1^2-\lambda_2^2} \la{alpha} \en
with $\sigma_1$ and $\sigma_2$ denoting the principal Cauchy stresses in the $x_1$- and $x_2$-directions. We make the assumption that both $\alpha$ and $\gamma$ are positive material constants, which is consistent with the Baker-Ericksen inequality $(\sigma_1-\sigma_2)(\lambda_1-\lambda_2)>0$ $(\lambda_1 \ne \lambda_2)$ \citep{be1954}.

Equation \rr{3.4} admits a \lq\lq normal mode" buckling solution of the form
\be \phi(r, z)=r J_1(\zeta r) S(\zeta z), \la{add1} \en
where $\zeta$ is a constant playing the role of wavenumber, $J_1(x)$ is Bessel's function of the first kind, and the other function $S(\zeta z)$ is to be determined. This is similar to looking for periodic bifurcation solutions proportional to ${\rm e}^{\ii k x_1}$ in rectangular coordinates. In fact, $J_1(x)$ is oscillatory although it also decays like $1/\sqrt{x}$ for large values of $x$ due to geometric spreading.

On substituting \rr{add1} into  \rr{3.4} and simplifying by making use of the identity
$$ J_\nu (x)= \frac{2 (\nu -1)}{x} J_{\nu-1}(x)-J_{\nu-2}(x), $$
we obtain
$
 J_1 (\zeta r ) \left\{ \gamma S^{(4)}(\zeta z)-2 \beta S''(\zeta z)+\alpha S (\zeta z)\right\}=0$.
It then follows that
\be \gamma S^{(4)}(\zeta z)-2 \beta S''(\zeta z)+\alpha S (\zeta z)=0, \la{add2} \en
which has the same form as in the plane-strain case.
Trying a solution of the form $S(x)={\rm e}^{x/\omega}$, we find that $\omega$ must satisfy the bi-quadratic equation
\be \alpha \omega^4-2 \beta \omega^2+\gamma =0, \;\;\;\; \Longrightarrow \;\; \omega^2=\frac{1}{\alpha} (\beta\pm \sqrt{\beta^2-\alpha \gamma}). \la{add3} \en
Thus, a general solution is given by
\be S(\zeta z) = C_1 \sinh \frac{\zeta}{\sqrt{k_1} } z+
C_2 \sinh \frac{\zeta}{\sqrt{k_2} } z+C_3 \cosh \frac{\zeta}{\sqrt{k_1} } z +C_4 \cosh \frac{\zeta}{\sqrt{k_2} } z, \la{3.15} \en
where
 \be k_1=\frac{1}{\alpha} (\beta-\sqrt{\beta^2-\alpha \gamma}), \;\;\;\;
 k_2=\frac{1}{\alpha} (\beta+\sqrt{\beta^2-\alpha \gamma}). \la{3.8} \en
The boundary conditions \rr{incre5a} take the form
\be
B_{2323} u_z+(B_{2332}+\bar{p}) v_r=0, \;\;\;\; {\rm on}\;\; z=\pm h/2, \la{3.16} \en
\be
  B_{1122} \frac{u}{r} +(B_{2222}+\bar{p}) v_z+B_{2233} u_r-p^*=0,\;\;\;\; {\rm on}\;\; z=\pm h/2. \la{3.17} \en
The  $p^*$ in \rr{3.17} can be eliminated by differentiating \rr{3.17} with respect to $r$ and then using \rr{incre1}$_1$ to eliminate $p^*_r$. We then obtain
 $$ (B_{1122}-B_{1111}-\bar{p}) \left( u_{rr}+\frac{u_r}{r}-\frac{u}{r^2}\right) $$
\be  -B_{2121} u_{zz}+(B_{2222}-B_{1221}-B_{1122}) v_{rz}=0, \;\;\;\; {\rm on}\;\; z=\pm h/2. \la{3.18} \en
On substituting \rr{add1} and \rr{3.15}  into the four boundary conditions \rr{3.16} and \rr{3.18}, we obtain four algebraic equations. By simple addition and subtraction, the four equations can be decoupled into two equations for $C_1$ and $C_2$ and another two equations for $C_3$ and $C_4$. Since the determinants of the two coefficient matrices cannot vanish simultaneously,  when the equations for $C_1$ and $C_2$ have non-trivial solutions, $C_3$ and $C_4$ must vanish. In this case, $\phi$, and hence the vertical displacement $v$, are odd functions of $z$. The associated modes are thus extensional. On the other hand, when the equations for $C_3$ and $C_4$ have non-trivial solutions, $C_1$ and $C_2$ must vanish, and the associated modes are flexural. The bifurcation conditions for extensional and flexural modes are thus given by
$$   \hspace{-5cm} \sqrt{k_1} \left(1 +   k_1\right) \left(k_2 (2 \beta
   +\gamma )-\gamma \right) \tanh \left(\frac{\zeta
   h}{2\sqrt{k_1}}\right)  $$   \be -\sqrt{k_2} \left(1 +
   k_2\right) \left(k_1 (2 \beta +\gamma )-\gamma \right) \tanh
   \left(\frac{\zeta  h}{2\sqrt{k_2}}\right)=0, \la{3.19} \en

$$   \hspace{-5cm} \sqrt{k_1} \left(1+k_1\right) \left(k_2
   (2 \beta +\gamma )-\gamma \right) \tanh \left(\frac{\zeta
   h}{2\sqrt{k_2}}\right)  $$ \be -\sqrt{k_2} \left(1+k_2\right) \left(k_1 (2 \beta +\gamma )-\gamma
   \right) \tanh \left(\frac{\zeta
   h}{2\sqrt{k_1}}\right)=0,\la{3.20} \en
   respectively.
Expanding \rr{3.19} to order $(\zeta h)^2$, we obtain
\be  \gamma  (\beta +\gamma )+ \frac{1}{24} \left\{\alpha  \gamma -(2 \beta +\gamma
   )^2\right\}( \zeta h )^2 +\cdots =0, \la{3.21} \en
   where we have used \rr{3.8} to eliminate $k_1$ and $k_2$.

From the discussion in \citet{fu2001}, we may postulate that the bifurcation condition for localized necking can be obtained by setting the leading order term in \rr{3.21} to zero, that is $ \beta +\gamma=0$ since $\gamma>0$, or equivalently,
\be B_{1111}+B_{2222}+2 B_{2121}-2 B_{1221}-2 B_{1122}=0. \la{bif} \en
In terms of the strain-energy function $W(\lambda_1, \lambda_2, \lambda_3)$, this bifurcation condition takes the form
\be
2 \lambda^2 W^{(0)}_2+W^{(0)}_{22}-2 \lambda^3W^{(0)}_{12}+\lambda^6W^{(0)}_{11}=0, \la{bif1} \en
where $W_2=\paa W/\paa \lambda_2$,  $W_{12}=\paa^2 W/\paa \lambda_1 \paa \lambda_2$ etc., and a superscript \lq\lq (0)" signifies evaluation at $(\lambda_1, \lambda_2, \lambda_3)=(\lambda, \lambda^{-2}, \lambda)$.

This bifurcation condition has the following interpretations. The nominal stresses in the $x_3$- and $x_2$-directions (i.e. $r$- and $z$-directions) are given by
$$ s_3=W_3-\lambda_3^{-1} \bar{p}, \;\;\;\; s_2=W_2-\lambda_2^{-1} \bar{p}. $$
On eliminating $\bar{p}$ using the condition $s_2=0$, we obtain
$$ s_3=W_3-\frac{\lambda_2}{\lambda_3} W_2=\frac{\paa}{\paa \lambda_3} W(\lambda_1, \lambda_1^{-1} \lambda_3^{-1}, \lambda_3), $$
where in the last term we are viewing $W$ as a function of $\lambda_1$ and $\lambda_3$.
Viewing $s_3$ as a function of $\lambda_1$ and $\lambda_3$, we obtain
 $$\left.\frac{\paa s_3}{\paa \lambda_3}\right|_{\lambda_1 {\rm fixed}}=W_{33} -\frac{1}{\lambda_1 \lambda_3^2 }   W_{32} +\frac{2}{\lambda_1 \lambda_3^3 } W_2-\frac{1}{\lambda_1 \lambda_3^2 } \left[ W_{32}- \frac{1}{\lambda_1 \lambda_3^2 }   W_{22}  \right].$$
 On evaluation at $(\lambda_1, \lambda_2, \lambda_3)=(\lambda, \lambda^{-2}, \lambda)$, we obtain
 \be \lambda^6 \left.\frac{\paa s_3}{\paa \lambda_3}\right|_{\lambda_1 {\rm fixed}}=\lambda^6 W^{(0)}_{33}-2\lambda^3 W^{(0)}_{32}+2 \lambda^2 W^{(0)}_2+W^{(0)}_{22}. \la{3.22} \en
 Since indexes $1$ and $3$ can be exchanged, this may be replaced by
 \be \lambda^6 \left.\frac{\paa s_3}{\paa \lambda_3}\right|_{\lambda_1 {\rm fixed}}=\lambda^6 W^{(0)}_{11}-2\lambda^3 W^{(0)}_{12}+2 \lambda^2 W^{(0)}_2+W^{(0)}_{22}. \la{3.23} \en
 It is seen that setting $\paa s_3/\paa \lambda_3=0$  gives the bifurcation condition \rr{bif}.

 In the case of localised bulging of an inflated rubber tube, the bifurcation condition was shown to be equivalent to the Jacobian determinant of the internal pressure $P$ and resultant axial force $N$ equal to zero when $P$ and $N$ are each viewed as functions of two deformation variables such as the internal volume and axial stretch \citep{fl2016}. In the current setting, it is natural to choose the nominal stresses $s_1$ and $s_3$ as the force variables and $\lambda_1$ and $\lambda_3$ as the deformation variables with $\lambda_2$ eliminated using the incompressibility condition $\lambda_2=\lambda_1^{-1} \lambda_3^{-1}$. Thus, we have
 \be s_1= \frac{\paa}{\paa \lambda_1} W(\lambda_1, \lambda_1^{-1} \lambda_3^{-1}, \lambda_3),\;\;\;\; s_3= \frac{\paa}{\paa \lambda_3} W(\lambda_1, \lambda_1^{-1} \lambda_3^{-1}, \lambda_3). \la{aug1} \en
 The associated Jacobian determinant can be evaluated as follows:
 \be
 \frac{\paa s_1}{\paa \lambda_1}\frac{\paa s_3}{\paa \lambda_3}-\frac{\paa s_1}{\paa \lambda_3}\frac{\paa s_3}{\paa \lambda_1}=
 \left(\frac{\paa s_3}{\paa \lambda_3}\right)^2-\left(\frac{\paa s_3}{\paa \lambda_1}\right)^2 =\left(\frac{\paa s_3}{\paa \lambda_3}+\frac{\paa s_3}{\paa \lambda_1}\right)\left(\frac{\paa s_3}{\paa \lambda_3}-\frac{\paa s_3}{\paa \lambda_1}\right), \la{add100} \en
 where all the partial derivatives are evaluated at $\lambda_1=\lambda_3=\lambda$ (and hence indexes $1$ and $3$ can be exchanged). Thus, the Jacobian determinant equal to zero implies
 \be \frac{\paa s_3}{\paa \lambda_3}-\frac{\paa s_3}{\paa \lambda_1}=0, \;\;\;\;{\rm or}\;\; \;\;\frac{\paa s_3}{\paa \lambda_3}+\frac{\paa s_3}{\paa \lambda_1}=0, \la{aug4} \en
 which may be compared with the condition $\paa s_3/\paa \lambda_3=0$ for necking.
 %
%
The two equations in \rr{aug4} are obviously equivalent to
 \be \left\{\lim_{\lambda_3 \to \lambda_1} \frac{s_3-s_1}{\lambda_3-\lambda_1}\right\}_{\lambda_1=\lambda}=0, \;\;\;\;{\rm or}\;\;\;\;
 \frac{d}{d\lambda} s_3(\lambda, \lambda)=0, \la{aug3} \en
 which are simply the conditions for the TK and limiting-point instabilities, respectively. Thus, for the current problem, {\it the Jacobian determinant equal to zero is not the bifurcation condition for necking}.

 For the purpose of illustration, we consider the following class of strain-energy functions:
 \be
W=\frac{2\mu_1}{m_1^2} (\lambda_1^{m_1}+\lambda_2^{m_1}+\lambda_3^{m_1} -3 )+\frac{2\mu_2}{{m_2}^2} (\lambda_1^{m_2}+\lambda_2^{m_2}+\lambda_3^{m_2} -3 ), \la{energy_2terms}
\en
where $\mu_1, \mu_2$, $m_1, m_2$ are material constants. We then have
\be \left\{\frac{\paa s_1}{\paa \lambda_3}\right\}_{\lambda_1=\lambda_3=\lambda}=\left\{\frac{\paa s_3}{\paa \lambda_1}\right\}_{\lambda_1=\lambda_3=\lambda}=\frac{2}{\lambda^2} \left( \mu_1 \lambda^{-2 m_1}+\mu_2 \lambda^{-2 m_2} \right), \la{aug5} \en
which is positive provided both $\mu_1$ and $\mu_2$ are positive. Assuming that the latter is true,
it then follows from \rr{aug4} that necking would occur after the TK instability but before the limiting-point instability. For instance, for the case when $\mu_2=0, \; m_1=1/2$, we have
\be \lambda_{\rm TK}=2^{2/3}, \;\;\;\; \lambda_{\rm necking}= 3^{2/3}, \;\;\;\; \lambda_{\rm limiting}=4^{2/3}, \la{three} \en
where the three subscripts in $\lambda$ signify association with the TK, necking, and limiting-point instabilities, respectively.  On the other hand, when
$\mu_2=\mu_1/12, \; m_1=1/2, \; m_2=2$ with which the material exhibits moderate stiffening for large values of $\lambda$, we have
$$ \lambda_{\rm TK1}=1.698,\;\;\;\;\lambda_{\rm TK2}=7.803, \;\;\;\; \lambda_{\rm necking 1}= 2.318, $$ $$ \lambda_{\rm necking 2}= 7.487,  \;\;\;\; \lambda_{\rm limiting1}= 2.957,\;\;\;\; \lambda_{\rm limiting2} = 7.102, $$
where we have two bifurcation values for each of the three instabilities.
Finally, when $\mu_2=\mu_1/80, \; m_1=1/2, \; m_2=4$ which will be considered further in our numerical simulations later, we have
\be \lambda_{\rm TK1}=1.647,\;\;\;\;\lambda_{\rm TK2}=3.403, \;\;\;\; \lambda_{\rm necking 1}= 2.439,\;\;\;\; \lambda_{\rm necking 2}= 2.919,  \;\;\;\; \lambda_{\rm limiting} \to \infty, \la{three1} \en
where we have two bifurcation values for both the TK instability and necking, but no limit points exist due to the stronger stiffening behaviour of the material model. To be more precise, the second bifurcation value for necking is really the value for {\it bulging} when unloading takes place.
We expect, however, that the TK instability can only occur under tension control even if the bifurcation condition \rr{aug4}$_1$ is satisfied. Under displacement control, the two in-plane principal stretches are forced to be equal, and we expect that the TK instability should give way to necking if \rr{bif1} is satisfied. These facts will indeed be confirmed by our numerical simulations to be conducted in Section 5.

 %
%
%
%
%
%


\section{Weakly nonlinear analysis}
\setcounter{equation}{0}
Having determined the bifurcation condition for necking, we now denote the critical stretch by $\lambda_{\rm cr}$ and conduct a weakly nonlinear analysis to derive an amplitude equation for the necking solution. To simplify notation, we scale $r$, $z$ and all displacement components by $h$, and use the same notations for the scaled quantities. As a result, we have $h=1$.

We use $\lambda_1=\lambda_3 \equiv \lambda$ as the control parameter in our near-critical analysis.
Guided by the analysis in Fu (2001), we may write
\be \lambda=\lambda_{\rm cr}+ \ep \lambda_0, \la{lambda0} \en
and define a far distance variable $s$ through
\be s=  \sqrt{\ep} r, \la{farz} \en
where $\lambda_0$ is a  constant and $\ep$ is a small positive parameter characterizing the order of deviation of $\lambda$ from its critical value $\lambda_ {\rm cr}$.  Note that as a result of \rr{lambda0},  the  moduli $B_{jilk}$ and $B^2_{jilknm}$ must all be expanded in terms of $\ep$ as well, but these expansions are not written out for the sake of brevity. The pressure $\bar{p}$ is eliminated using the identity $\bar{p}=B_{2121}-B_{2112}$.

By analyzing the linear results in Section 3 in the limit $\lambda \to \lambda_{\rm cr}$, we may deduce that
$ v=O(\sqrt{\ep} u),  p^*=O(\sqrt{\ep} u)$. From the fact that the $\lambda_0$ in \rr{lambda0} must appear at the order where a solvability condition should be imposed, we find that
 $u$ must be of order $\sqrt{\ep}$. Thus, we look for an asymptotic solution of the form
$$   u=\sqrt{\ep} \left\{ u^{(1)}(s, z)+\ep u^{(2)}(s, z)+\ep^2 u^{(3)}(s, z)+\cdots \right\}, $$
\be v= {\ep} \left\{ v^{(1)}(s, z)+\ep v^{(2)}(s, z)+\ep^2 v^{(3)}(s, z)+\cdots \right\}, \la{asymsol} \en
$$   p^*=\ep \left\{ p^{(1)}(s, z)+\ep p^{(2)}(s, z)+\ep^2 p^{(3)}(s, z)+\cdots \right\}, $$
where all the functions on the right hand sides are to be determined from successive approximations.

On substituting \rr{asymsol} into the equilibrium equations \rr{incre1}, the incompressibility condition \rr{incre4}, the boundary conditions \rr{incre5a}, and then equating the coefficients of like powers of $\ep$, we obtain a hierarchy of boundary value problems. The solutions at different orders are obtained as follows. In the following description, the two equilibrium equations in \rr{incre1} are referred to as the $r$- and $z$-equilibrium equations, respectively.

At leading order, the $r$-equilibrium equation subject to the second boundary condition in \rr{incre5a} yields
\be
u^{(1)}(s, z)=A(s), \la{4.5} \en
where $A(s)$ is to be determined. Integrating the incompressibility condition then gives
\be
v^{(1)}(s, z)=-z \frac{1}{s} (s A(s))'+B(s), \la{4.6} \en
where $B(s)$ is another function to be determined. Integrating the $z$-equilibrium equation subject to the first boundary condition at $z=\pm h/2$ yields the following  expression for $p^{(1)}(s, z)$:
\be
p^{(1)}(s, z)=-(B_{2222}-B_{1122}+B_{2121}-B_{2112}) \frac{1}{s} (s A(s))', \la{4.7} \en
where here and hereafter in this section all the moduli are evaluated at $\lambda=\lambda_{\rm cr}$.

At second order, solving the $r$-equilibrium equation yields an expression for $u^{(2)}(s, z)$ that contains two new functions $C(s)$ and $D(s)$ in the form $C(s)+z D(s)$.  Subtracting and adding the second boundary condition at $z=\pm h/2$, respectively, we obtain
\be
B_{2222}-2 B_{1122}+B_{1111}+2 B_{2121}-2 B_{2112}=0, \la{4.8} \en
and
\be D(s)=- B'(s). \la{4.9} \en
The first result \rr{4.8}  may be shown to be equivalent to the bifurcation condition \rr{bif1}.
Integrating the incompressibility condition at second order gives an expression for $v^{(2)}(s, z)$ that contains a new function $F(s)$, whereas integrating the $z$-equilibrium yields an expression for $p_2(s, z)$ containing a new function $E(s)$. Finally,  subtracting and adding the second boundary condition at $z=\pm h/2$, respectively, we obtain $s B''(s)+B'(s)=0$, and an expression for E(s). Solving the equation for $B(s)$, we obtain  $B(s)=d_1 \ln s+d_2$. Since $v^{(1)}$ should be bounded at $s=0$, and without loss of generality we may also impose the condition $v^{(1)}(0,0)=0$, we must set $d_1=d_2=0$. Thus we may set $ B(s)=0$. This then completes the solution at second order.

At third order, the nonlinear terms come into play and a solvability condition must be imposed. The $r$-equilibrium equation can be solved to find an expression for $u^{(3)}(s, z)$ that contains two new functions $G(s)$ and $H(s)$ in the form $G(s)+z H(s)$. Subtracting and adding the second boundary condition evaluated at $z=\pm h/2$, respectively, we obtain the amplitude equation for $A(s)$ and an expression for $H(s)$. 
After some simplification, it is found that the amplitude equation takes the form
\be c_0 \frac{d}{ds} \frac{1}{s} \frac{d}{ds} s P'(s)+ c_1 \lambda_0 P'(s)+c_2 \frac{d}{ds} P^2(s)+
c_3 A''(s) \left( A'(s)-\frac{1}{s} A(s)\right)=0, \la{4.10} \en where a prime signifies differentiation, $P(s)$ is defined by
\be P(s)=\frac{1}{s} (s A(s))', \la{4.11s} \en
and the three coefficients are given by
\begin{eqnarray}
  c_0&=&\frac{1}{12} \left(B_{1212}-B_{2121}\right), \nonumber \\
 c_1&=&2 B'_{1122}+2 B'_{1221}-B'_{1111}-2
   B'_{2121}-B'_{2222}, \nonumber \\
 c_2 &=& \frac{1}{4} \left(-4 B_{1111}-2 B_{1122}+6 B_{2222}-B^2_{111111}+4
   B^2_{111122}-B^2_{111133} \right. \nonumber \\
   & & \left.-6 B^2_{112222}+2 B^2_{112233}+2
   B^2_{222222}\right), \nonumber \\
c_3&=&B_{1122} -B_{1111}+B^2_{111122}-B^2_{112233}-\frac{1}{2}B^2_{111111}+\frac{1}{2}B^2_{11
   1133}. \nonumber \end{eqnarray}
In the above expressions, $B'_{1122}$ denotes $d B_{1122}/d\lambda$ etc., and we have used the bifurcation condition \rr{4.8} to eliminate $B_{2112}$.  When the strain energy function is given by \rr{energy_2terms} with $\mu_2=0$, these constants take the form
$$ c_0=\frac{\lambda^{-2 m_1} \left(2 m_1 (\lambda_{\rm cr}^6+1)
    (\lambda_{\rm cr}^{6 m_1}-1)-m_1^2 (\lambda_{\rm cr} ^6-1)  (\lambda_{\rm cr}^{3 m_1}+1)^2-(\lambda_{\rm cr} ^6-1) (\lambda_{\rm cr}^{3 m_1}-1)^2\right)}{24 m_1 (\lambda_{\rm cr} ^{3 m_1}-1)}, $$
   $$ c_1=2 \lambda_{\rm cr}^{-2 m_1-1} \left(\lambda_{\rm cr}^{3 m_1}-m_1 (\lambda_{\rm cr}^{3 m_1}-2)+2\right),\;\;\;\;c_3=-(m_1-1) \lambda_{\rm cr}^{m_1}, $$ $$
c_2=-\frac{(m_1-1)  \left(m_1 \left(\lambda_{\rm cr}^{3 m_1}-2\right)+2 \left(\lambda_{\rm cr}^{3 m_1}-1\right)\right)}{2 m_1 \lambda_{\rm cr}^{2 m_1}}, $$
where $\lambda_{\rm cr}=\left\{(1+m)/(1-m)\right\}^{1/(3m)}$. More specifically when $m_1=1/2$, we have
$$ c_1/c_0= 9 \sqrt[3]{3}/2, \;\;\;\; c_2/c_0=27/8, \;\;\;\; c_3/c_0=9/4. $$
As a consistency check, we may neglect the nonlinear terms in \rr{4.10} to obtain
\be c_0 \frac{d}{ds} \frac{1}{s} \frac{d}{ds} s P'(s)+ c_1 \lambda_0 P'(s)=0. \la{may1} \en
On substituting a solution of the form $A(s)=J_1(k s)$ into \rr{4.11s} and the resulting expression into \rr{may1} and simplifying, where $k$ is a constant, we obtain
\be c_1 \lambda_0 -c_0 k^2 =0. \la{may2} \en
On the other hand, expanding \rr{3.21} around $\lambda=\lambda_{\rm cr}$ and taking $\zeta=k \sqrt{\ep}, h=1$, we obtain
\be  \left\{\frac{d}{d\lambda}\gamma  (\beta +\gamma )\right\}_{\rm cr} \lambda_0 + \frac{1}{24} \left\{\alpha  \gamma -(2 \beta +\gamma
   )^2\right\}_{\rm cr}    k^2=0, \la{may3} \en
where the subscripts \lq\lq cr" signify evaluation at $\lambda=\lambda_{\rm cr}$. We have verified that \rr{may2} is indeed consistent with \rr{may3}.

As another consistency check, we may expand \rr{4.11s} fully out and omit all the terms that are divided by powers of $s$ to obtain its planar counterpart:
\be
c_0 A^{(4)}(s)+c_1 \lambda_0 A''(s)+ c_2^* A'(s) A''(s)=0, \la{2damp} \en
where
\be c_2^*=2 c_2+c_3=3 B_{2222}-3 B_{1111}-B^2_{111111}+3 B^2_{111122}-3
   B^2_{112222}+B^2_{222222}. \la{2dc2} \en
Equation \rr{2damp} applies to a finitely deformed sheet that is subjected to plane-strain incremental deformations, and recovers the static counterpart of the amplitude equation (4.41) in \citet{fu2001}. It has an exact solution given by
\be A(s)=\frac{6 c_0}{c_2^*} \sqrt{\frac{-c_1 \lambda_0}{c_0}} {\rm tanh} \left( \frac{1}{2} \sqrt{\frac{-c_1 \lambda_0}{c_0}} s\right). \la{plane} \en
This solution has the property $A'(s) \to 0$ as $s \to \infty$ and is the localised necking solution in the 2D case.

Although the solution \rr{plane} for the plane-strain case tends to a constant as $s \to \infty$, we expect its counterpart for the axisymmetric case to decay algebraically due to geometrical spreading. Thus, for large $s$, we may neglect the quadratic part in \rr{4.10} to obtain
\be   \frac{d}{ds} \frac{1}{s} \frac{d}{ds} s P'(s)-\kappa^2 P'(s)=0, \la{4.10f} \en
where $\kappa^2=- c_1 \lambda_0/c_0$.
Integrating once and setting the constant to zero, we obtain
$$ s^2 P''(s)+s P'(s)-\kappa^2 s^2 P(s)=0. $$
This is a modified Bessel's equation. It has a decaying solution given by
\be P(s)= a_1 K_0( \kappa s), \la{march2} \en
where $K_0$ is the modified Bessel function of the second kind and $a_1$ is a constant. On substituting \rr{march2} into \rr{4.11s} and integrating, we obtain
\be A(s)=A_\infty(s) \equiv \frac{a_2}{s}-\frac{a_1}{\kappa} K_1( \kappa s), \la{march3} \en
where $a_1$ and $s_2$ are constants, and $K_1$ is the modified Bessel function of the second kind that has the asymptotic behaviour
\be K_1(x) \sim  \sqrt{\frac{\pi}{2 x}}  \left[1+\frac{3}{8x}+\cdots \right] \,{\rm e}^{-x}, \;\;\;\;{\rm as}\;\; x \to \infty. \la{k1} \en
The asymptotic behaviour \rr{march3} is consistent with our earlier assumption that $A(s)$ decays algebraically. In principle, the 4th-order differential equation \rr{4.10} can be integrated subject to the conditions
\be A(s), A''(s) \to 0 \;\; {\rm as}\; s \to 0, \la{bc1} \en
and
\be A(s) \to  A_\infty(s), \;\; A'(s) \to A_\infty'(s) \;\;\;{\rm as}\; s \to \infty. \la{bc2} \en
The necessary numerical integration will not be carried out here. Instead, we shall proceed to a fully nonlinear numerical simulation of the necking solution in the next section. We conclude this section by observing that the decay behaviour \rr{march3} requires $\kappa$ ($=\sqrt{- c_1 \lambda_0/c_0}$) to be real. For all the material models considered, it is found that
$c_1/c_0>0$. Thus, bifurcation into a localised necking solution is generally sub-critical (i.e. $ \lambda_0<0$).

\section{Fully nonlinear numerical simulations}
\setcounter{equation}{0}
To verify the various theoretical predictions presented in the previous sections, we now conduct nonlinear finite element (FE) simulations within Abaqus 2020/Standard \citep{ab2013} to investigate the effect of loading methods and material properties on the type of bifurcation and post-bifurcation behavior. In particular, we shall verify that when the loading is force controlled, TK instability will precede necking, whereas when the loading is displacement controlled, TK instability will be suppressed and it is necking that will occur.

\subsection*{Equibiaxial stretching of a square plate}
\begin{figure*}[h!]
\centering
\includegraphics[width=1.0\linewidth]{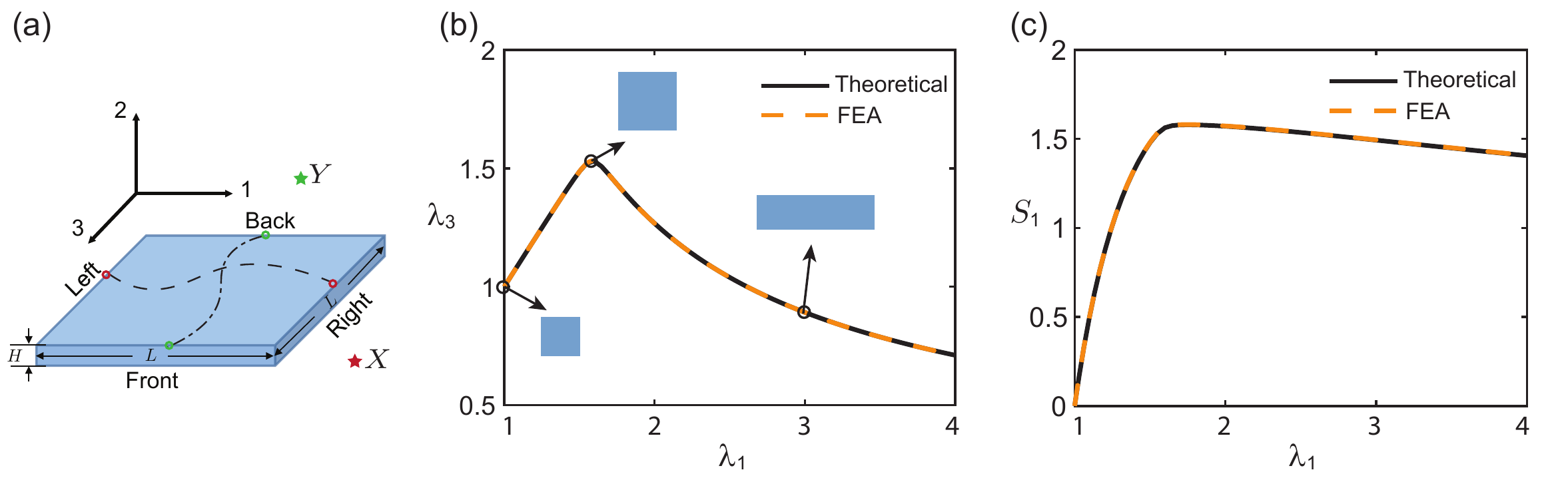}
\caption{Comparison between numerical results and theoretical predictions for the Treloar-Kearsley instability. (a) Geometrical parameters and boundary conditions for the square plate. (b) $\lambda_3$ versus $\lambda_1$. (c) Nominal stress $S_1$ scaled by $\mu_1$ versus $\lambda_1$. }
\label{fig_tk_simulation}
\end{figure*}

We first perform fully 3D FE simulations for a square plate with $L=5H$, where $L$ is the length and width of the plate and $H$ is the thickness of the plate (Figure~\ref{fig_tk_simulation}a). The plate is meshed using 8-node fully integrated hybrid linear brick elements (Abaqus Element C3D8H).
The following constraints are imposed on the opposite faces of the plate
\be u_\alpha^{L_i}=u_\alpha^{R_i}+u_\alpha^X,\;\;\;\;
     u_\alpha^{F_i}=u_\alpha^{B_i}+u_\alpha^Y, \;\;\;\;(\alpha=1, 2, 3, \;\; i=1,2,...,N),\la{eq-5-pbc} \en
where $N$ is the number of nodes on each of the four faces (with $L$, $R$, $F$ and $B$ denoting left, right, front and back faces, respectively), and $u_\alpha^{L_i}$ denotes the displacement in the $\alpha$-direction at the $i$-th node of the left face etc.
The $u_\alpha^X$ and $u_\alpha^Y$ denote the displacements in the $\alpha$-direction at two arbitrarily chosen reference points $X$ and $Y$ that are used to apply the loading. In the above notations,
the paired nodes  $L_i$ and $R_i$ have the same coordinates in the \textit{2-3 plane}, whereas $F_i$ and $B_i$ have the same coordinates in \textit{1-2 plane}. Moreover, to prevent rigid body motions, we fix the node at the center of the plate.

 The plate is loaded by applying the following concentrated forces and displacement constraints at the two reference points:
\begin{align}\la{eq-5-ref}
\nonumber
    & f_1^X=1.001*\bar{f}, ~~~u_2^X=u_3^X=0, ~~\text{at point $X$},\\
    & f_3^Y=\bar{f}, ~~~~~~~~~~~~~u_1^Y=u_2^Y=0, ~~\text{at point $Y$},
\end{align}
where $f_1^X$ and $f_3^Y$ are the concentrated forces applied at the reference points $X$ and $Y$ in the $1$- and $3$-directions, respectively, and $\bar{f}$ represents the magnitude of the concentrated force. The factor $1.001$ in $f_1^X$ can be viewed as an imperfection that will trigger the TK instability.
Note that the near equibiaxial conditions can also be obtained by applying near equal pressures directly at the edges of the plate, but this will result in convergent issues in simulations. The imposed constraints \rr{eq-5-pbc} serve to circumvent convergent issues and ensure a homogeneous deformation.

It is known that the bifurcation associated with the TK instability can be either sub-critical or super-critical depending on the strain-energy function used. The corresponding test has been given by \citet{og1987} in his equation (3.13). We thus consider two representative material models exhibiting these different behaviors. The first model is given by \rr{energy_2terms} with $\mu_1=1, \mu_2=0$ and $m_1= 1/2$ for which the bifurcation is sub-critical. We employ the modified RIKS algorithm \citep{riks1979} to capture the unloading behavior associated with the sub-critical bifurcation. We find that, initially, the common principal stretch in the $1$- and $3$-directions increases uniformly as the load increases. Then, when $\lambda_1$ reaches the critical value $1.58$ given by \rr{three}, bifurcation occurs, resulting in unequal stretches and decline of the nominal stress in both directions. As Figure~\ref{fig_tk_simulation}(b, c) shows, there is very good agreement with the theoretical predictions for both the stretches and nominal stress.

\begin{figure*}[ht]
\centering
\includegraphics[width=0.5\linewidth]{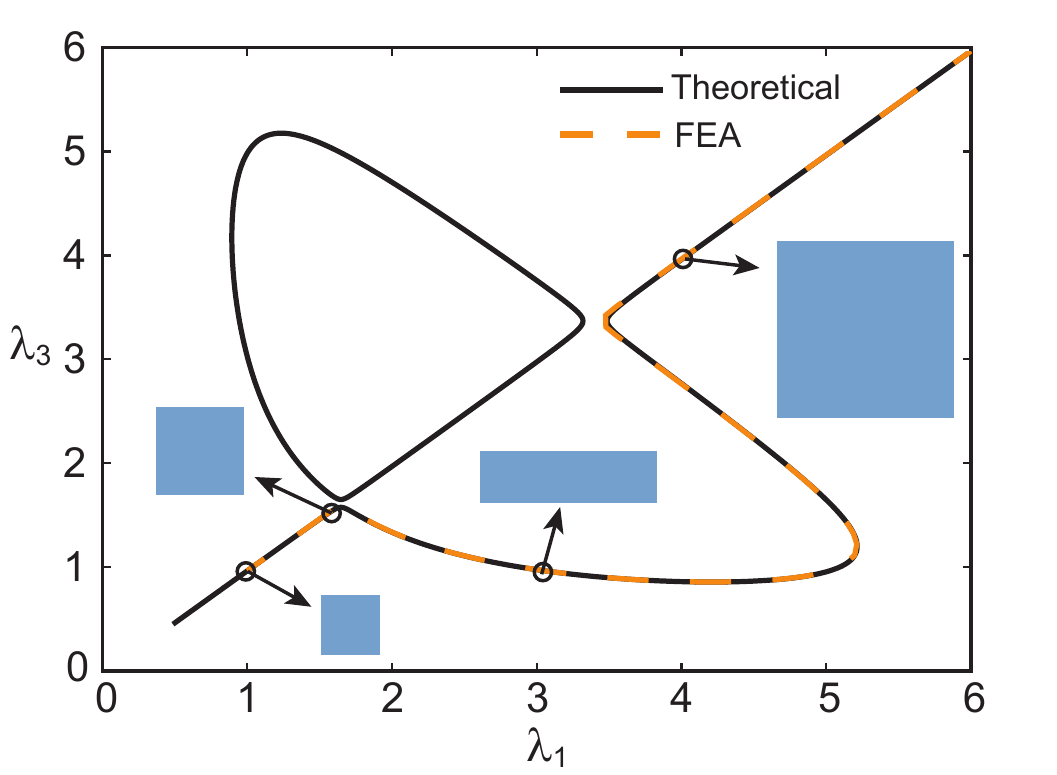}
\caption{Emergence and disappearance of the Treloar-Kearsley instability for the material model \rr{energy_2terms} with $\mu_2/\mu_1=1/80,\, m_1= 1/2$, $m_2=4$.}
\label{fig_tk_simulation_2_terms}
\end{figure*}

The second material model that we consider is given by \rr{energy_2terms} with $m_1= 1/2$, $m_2=4$, and $0.012<\mu_2/\mu_1<0.035$. The two limits of $\mu_2/\mu_1$ are such that if $\mu_2<0.012$ the bifurcation will be sub-critical, whereas if the upper limit is exceeded TK instability can no longer occur. The second term in \rr{energy_2terms} introduces straining-stiffening effect, and as a result there are two bifurcation stretches associated with the TK instability. It is observed that the nonuniform stretches in the $1$- and $3$-directions initiating from the first bifurcation stretch would become uniform again at the second bifurcation stretch. This is demonstrated in Figure \ref{fig_tk_simulation_2_terms} where we compare the numerical and theoretical results for the principal stretch $\lambda_3$ as a function of $\lambda_1$ when $\mu_2/\mu_1=1/80$.  Again, we have excellent agreement between numerical results and theoretical predictions.

\begin{figure*}[ht]
\centering
\includegraphics[width=0.6\linewidth]{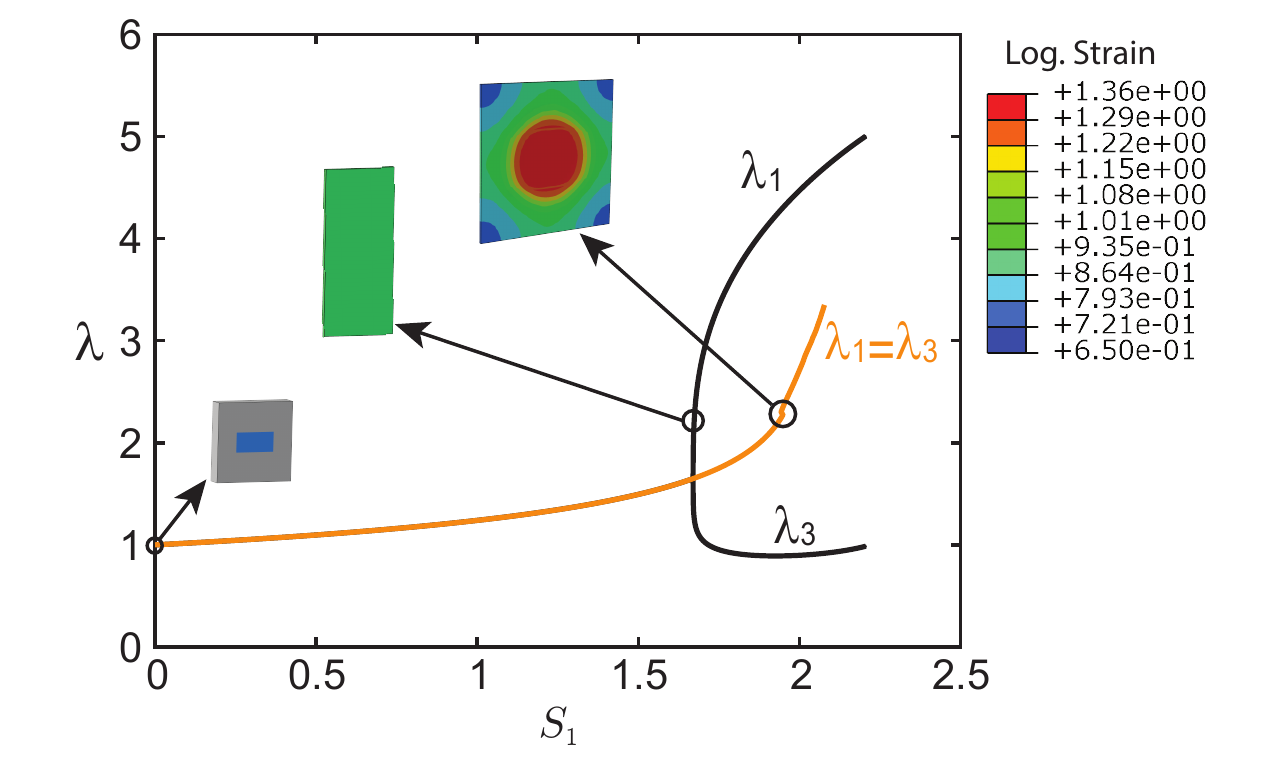}
\caption{Numerical simulations of a square plate with  material imperfections, modeled by the strain-energy function \rr{energy_2terms} with $m_1=1/2$, $m_2=4$ and $\mu_2/\mu_1=1/80$.  Solid yellow line: results obtained by applying equal displacements at the reference points using the STATIC module in Abaqus (in which case TK instability is suppressed and necking takes place at the kinked point in the black circle). Solid black line: results obtained by applying equal concentrated forces (force control) at the reference points using the RIKS module in Abaqus (in which case the TK instability occurs as a super-critical bifurcation). }
\label{fig_tk_necking}
\end{figure*}

Moreover, we remark that the TK instability can also be observed in an equibiaxially stretched square plate by introducing material imperfections. To demonstrate this, we introduce a small  imperfection  at the center of the plate by assuming that the shear modulus in a small {\it rectangular} area is 0.5\% smaller than that for the surrounding material; see the schematic in Figure \ref{fig_tk_necking} where the blue region in the square plate has the modulus reduced. Then, by applying equal concentrated forces at the reference points X and Y, we can observe unequal stretches when the applied nominal stress $S_1$ is larger than 1.67 (see the solid black lines in Figure \ref{fig_tk_necking}, which are obtained by RIKS simulation in Abaqus). However, if the reference points are stretched using displacement control (i.e. by applying equal displacements in both directions), TK instability will be suppressed, and instead, localized necking will appear at a larger  stretch as indicated by the solid yellow line in Figure \ref{fig_tk_necking}, which is obtained using STATIC simulation in Abaqus. The TK instability giving way to necking under displacement control will be discussed further in the next subsection. In addition, we note that a rectangular (anisometric) imperfection is essential for observing the TK instability in a square plate. If a square (isometric) defect were introduced instead, the TK instability would not be triggered even with force control.

\subsection*{All-round stretching of a circular plate}
\begin{figure*}[ht]
\centering
\includegraphics[width=0.9\linewidth]{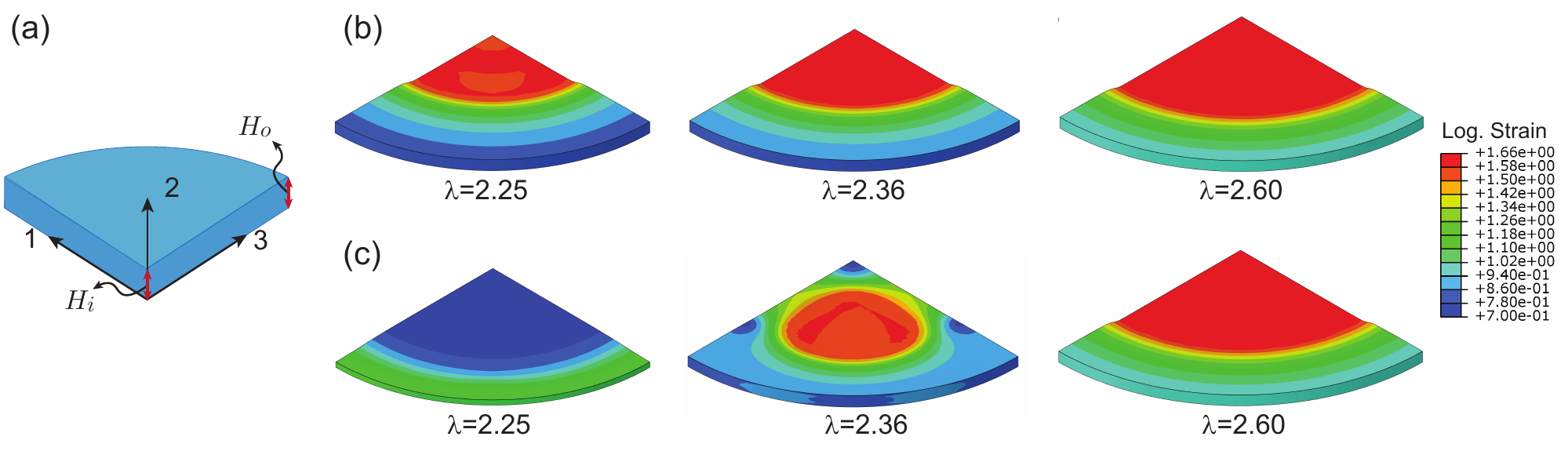}
\caption{Simulations of necking in a circular plate subject to an all-round stretch, modeled by the strain-energy function \rr{energy_2terms} with $m_1=1/2$, $m_2=4$ and $\mu_2/\mu_1=1/120$. (a) Geometric parameters for a quarter of the plate. (b,c) Numerical snapshots of the plate subjected to increased stretches with gradual thickness thinning towards the center (b) or the edge (c). }
\label{fig_postbifurcation_FE}
\end{figure*}

To demonstrate the occurrence of necking in a more convenient way than in the above square plate setting,
we now consider a circular plate with $R_0=2 H$, where $R_0$ and $H$ are respectively the radius and thickness of the plate in the undeformed configuration. We control the amount of stretch on the circular edge of the plate in order to suppress the occurrence of TK instability. In addition, since strain-stiffening is essential for observing the stable propagation stage of necking, we shall consider the two-terms material model~(\ref{energy_2terms})
with $\mu_2 \ne 0, \; m_2 \ge 2$. To reduce the cost of simulations, we only model a quarter of the plate and apply axisymmetric boundary conditions on the symmetric surfaces shown in Figure~\ref{fig_postbifurcation_FE}a.  Moreover, we note that the initiation location of necking is sensitive to geometric imperfections. In Figure~\ref{fig_postbifurcation_FE}b and c, we showcase snapshots of our simulations with gradual thickness thinning towards the origin ($H_i=0.9999~H_o$) or the circular edge ($H_o=0.9999~H_i$), respectively, where $H_i$ and $H_o$ denote the plate thicknesses at the center and outer edge.
For the case with thinning towards the origin, necking initiates at the center of the plate and propagates outwards as the applied load increases (Figure~\ref{fig_postbifurcation_FE}(b)). In contrast, when thinning towards the outer edge is introduced, necking initiates at the outer edge and spreads towards the center as the applied load increases. At some stage, however, the necked region switches to the center of the plate and propagates outwards with increased load (Figure~\ref{fig_postbifurcation_FE}(c)). Thus, necking starting at the center seems to be the dominant deformation mode, and so from now on we only consider plates with thinning towards the origin. In addition, the critical stretch for necking initiation is sensitive to the magnitude of geometric imperfections. For consistence, all the following numerical results correspond to the same geometric imperfection mentioned above. Due to the presence of near-critical snap-through behavior, we also add volume-proportional damping to the model to facilitate convergence (using the option STABILIZE in Abaqus), and set the dissipated energy fraction equal to ${\rm e}^{-10}$ and the maximum ratio of stabilization to strain energy equal to 0.01. Furthermore, we observe that although 2D axisymmetric simulations in Abaqus considering only a rectangular slice can reduce the cost of simulations significantly, the 2D model does not seem to be capable of characterizing the necking phenomenon in the sense that axisymmetric elements are prone to distortion when necking occurs, especially for materials with only a moderate strain-stiffening behavior.

\begin{figure*}[ht]
\centering
\includegraphics[width=0.4\linewidth]{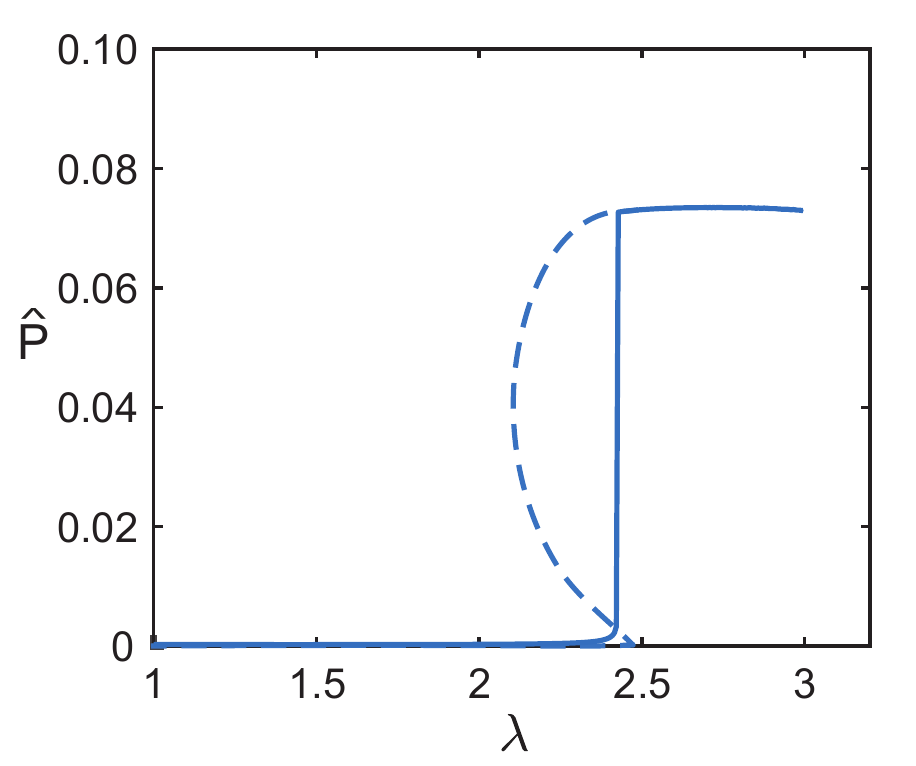}
\caption{Dependence of the necking amplitude on the applied stretch $\lambda$. The solid curve represents numerical results for the amplitude evolution at the center of the plate. The dashed line is a sketch of how the actual solution should look like. The material is modeled by the strain-energy function \rr{energy_2terms} with $m_1=1/2$, $m_2=4$ and $\mu_2/\mu_1=1/80$.}
\label{fig_necking_postbifurcation88}
\end{figure*}

\begin{figure*}[ht]
\centering
\includegraphics[width=0.75\linewidth]{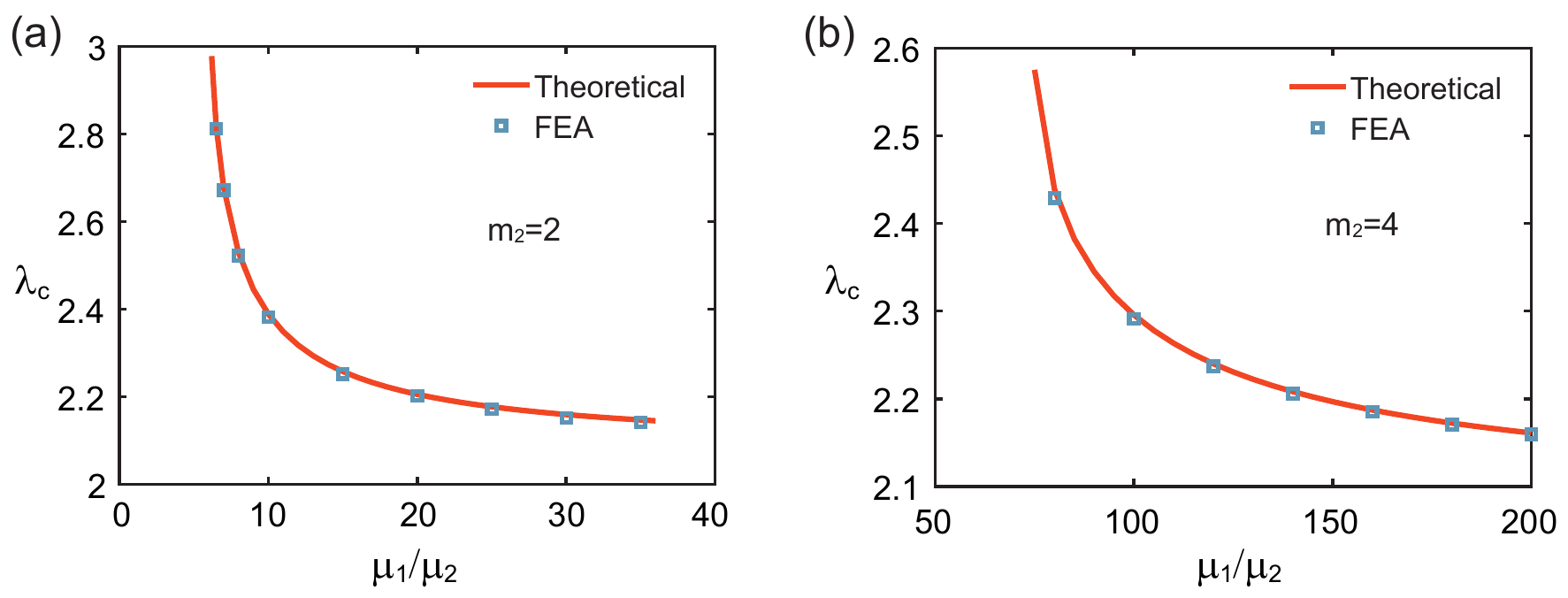}
\caption{Dependence of the critical stretch on $\mu_1/\mu_2$ with (a) $m_1=1/2$ and $m_2=2$ or  (b) $m_1=1/2$ and $m_2=4$.}
\label{fig_necking_bifurcation}
\end{figure*}

Figure~\ref{fig_necking_postbifurcation88} shows a typical plot of the necking amplitude $\hat{P}$ against the stretch $\lambda$, where $\hat{P}$ denotes the
absolute difference between the vertical displacements at points $(0, -H_i/2, 0)$ and  $(R_0, -H_i/2, 0)$ normalised by $H_i$.
It is seen that it is appropriate to take the numerical value of critical stretch as the stretch at which an amplitude jump takes place (the snap-through behavior mentioned above). We observe, however, that the variation of amplitude against stretch should have a slight snap-back (indicated by the dashed line) that is not captured by our simulations because the snap-back section corresponds to an unstable solution; see Figure 3(b) in \citet{fx2010} for a discussion of the same phenomenon in the context of inflation of a membrane tube.
To verify our theoretical predictions, we fix $m_1=1/2$, and compare the numerical and theoretical results in Figure~\ref{fig_necking_bifurcation} for the critical stretch as a function of $\mu_1/\mu_2$ for two typical values of $m_2$. It is seen that for each given $m_2$, the critical stretch is a monotonically decreasing function of the ratio $\mu_1/\mu_2$, and the excellent agreement between the numerical and theoretical results confirms that the bifurcation condition for necking is indeed given by \rr{bif}, not by the criterion based on the Jacobian determinant.

\begin{figure*}[ht]
\centering
\includegraphics[width=0.8\linewidth]{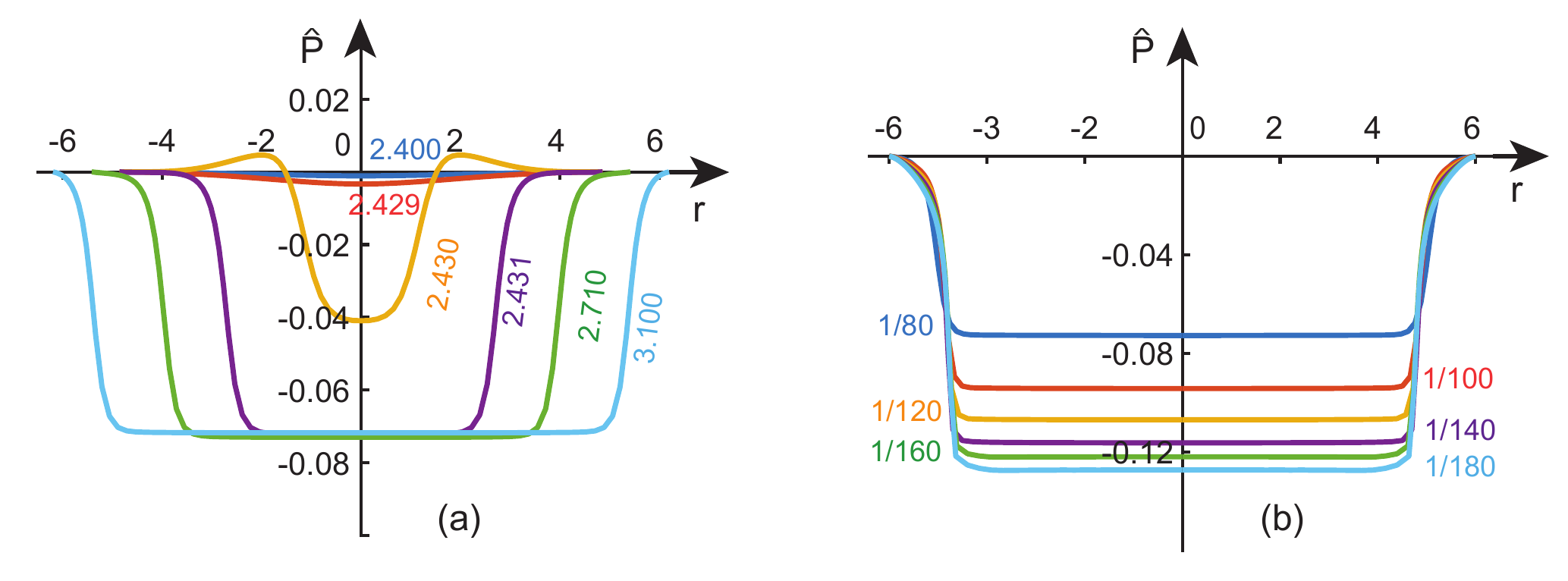}
\caption{ (a) Evolution of the necking profile with $m_1=1/2$, $m_2=4$, $\mu_2/\mu_1=1/80$. The six profiles correspond to $\lambda=2.40,~2.429,~2.430,~2.431,~2.710$ and $3.10$, respectively. (b) Evolution of the necking amplitude of the plate with $m_1=1/2$, and $m_2=4$ when subjected to equal-biaxial stretch of $\lambda=3.0$. The six curves represent $\mu_2/\mu_1=1/80,~1/100,~1/120,~1/140,~ 1/160$ and $1/180$, respectively.}
\label{fig_necking_postbifurcation}
\end{figure*}

Furthermore, to quantify the post-bifurcation behavior of the necking deformation, we report evolution of the necking amplitude in Figure \ref{fig_necking_postbifurcation}. The results shown in Figure \ref{fig_necking_postbifurcation}(a) represent a plate with $m_1=1/2$, $m_2=4$, $\mu_2/\mu_1=1/80$, and the six profiles correspond to $\lambda=2.40$, 2.429, 2.430, 2.431, 2.710 and 3.10, respectively (the critical stretch being 2.439).
The first two profiles correspond to tiny inhomogeneous deformations due to thickness thinning, whereas the third profile with $\lambda=2.430$ corresponds to the profile
immediately after the jump in Figure~\ref{fig_necking_postbifurcation88} has taken place. Thus, the critical stretch found numerically is $2.430$ which differs from the exact critical stretch $2.439$ given by \rr{three1}$_2$ with a relative error less than 0.4\%. It is observed that the necking amplitude first increases rapidly with respect to $\lambda$, and then as a maximum amplitude is approached the necked section begins to propagate outwards. This is reminiscent of the well-known growth/propagation behavior of localised bulges in an inflated rubber tube \citep{kc1991, wg2019}.
To demonstrate the effect of strain-stiffening on the necking amplitude, we report in Figure \ref{fig_necking_postbifurcation}(b) the dependence on $\mu_2/\mu_1$ of the necking amplitude with  $m_1=1/2$, $m_2=4$ and a fixed stretch of $\lambda=3.0$.  It can be seen that the necking amplitude increases as the value of $\mu_2/\mu_1$ decreases, which implies that the necking amplitude is significantly affected by the strain-stiffening response of the material.
\begin{figure*}[ht]
\centering
\includegraphics[width=0.8\linewidth]{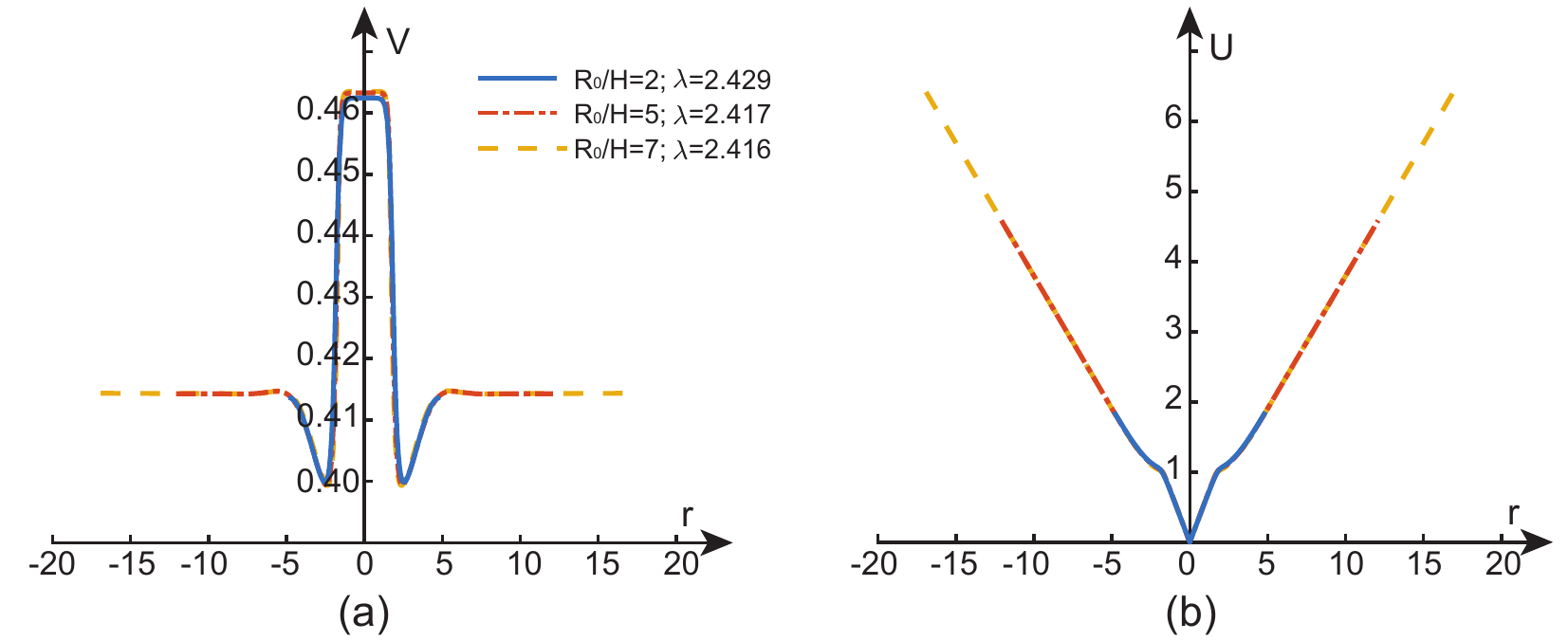}
\caption{ Total displacement (i.e. displacement measured from the undeformed configuration) in the (a) $z$-direction and (b) radial direction, at the lower surface of the plate as a function of the radial coordinate $r$ in the current configuration.}
\label{fig_convergence}
\end{figure*}

Finally, we remark that although the radius/thickness ratio used in our simulations is only 2 initially ($R_0/H=2$), this ratio has become as large as $7.62$ when the stretch reaches its critical value $2.439$. We have run simulations to verify that this choice of initial aspect ratio is sufficient for obtaining results that are almost free of edge effects. To demonstrate this, we have shown in Figure \ref{fig_convergence} two representative simulations corresponding to $R_0/H=5, 7$ superimposed on the results for $R_0/H=2$.  Since $\lambda$ is the deformed radius divided by the undeformed radius, and is no longer a true characterisation of local radial stretch after necking has taken place, we choose the three stretches shown in  \ref{fig_convergence}(a) such that the maxima of $V$ are the same (or closest numerically) and check whether the remaining profiles would then also coincide. It is seen that this is indeed the case for both $V$ and $U$ although the three profiles for each displacement terminate at different values of $r$. Notably, to demonstrate the negligible effect of radius/thickness ratio on the observed necking phenomenon with various imperfections, the results shown in Figure 8 correspond to a circular material imperfection whereby the shear modulus of the material within $R<H$ is 0.01\% smaller than that of the surrounding material.

\begin{figure*}[ht]
\centering
\includegraphics[width=0.8\linewidth]{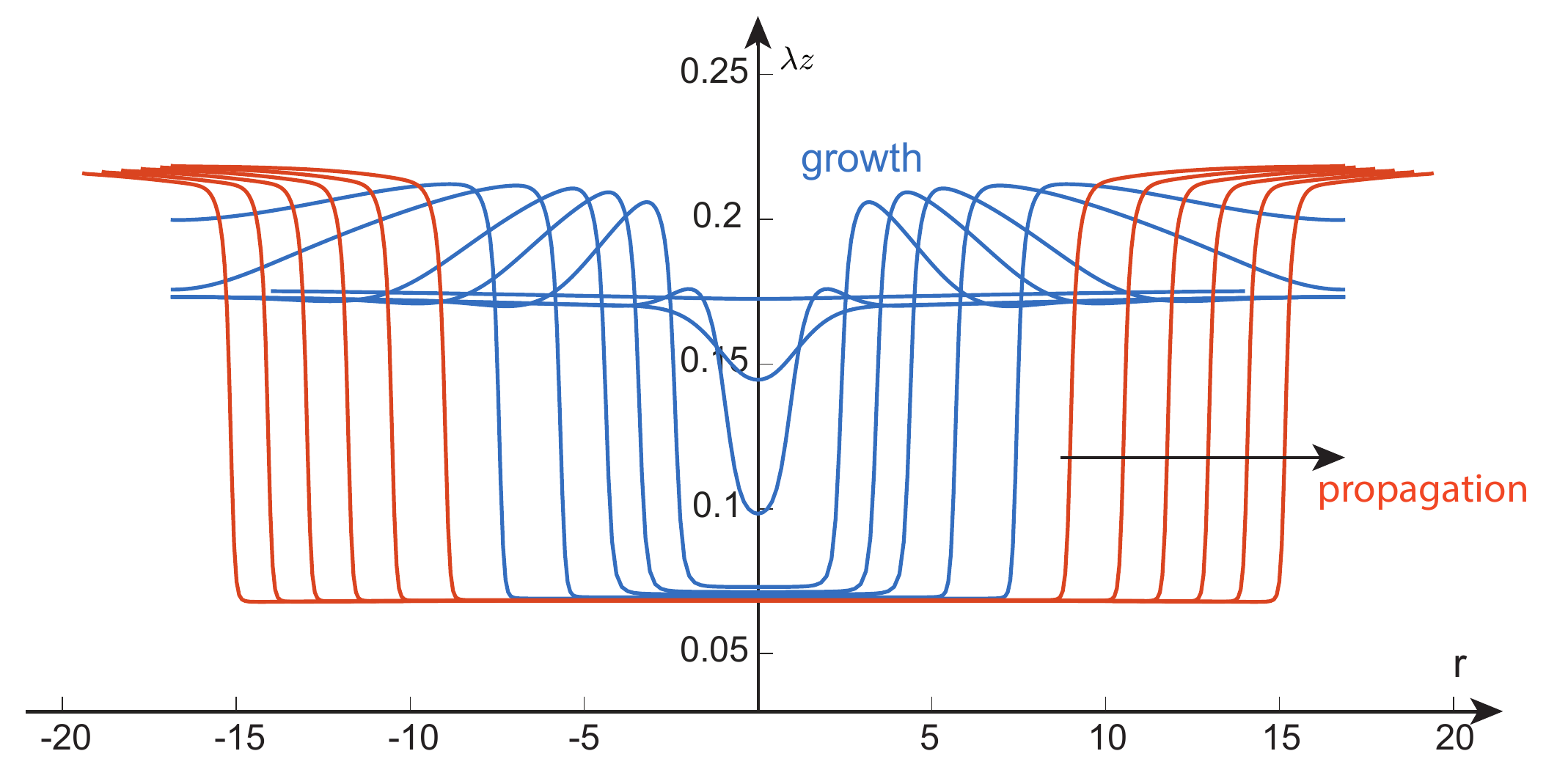}
\caption{Numerical simulation of the evolution of the principal stretch in the $z$-direction as the radius is increased. The strain energy function is given by \rr{energy_2terms} with $m_1=1/2$, $m_2=4$ and $\mu_2/\mu_1=1/80$, and the aspect ratio is $R_0/H=7$. The blue and red curves correspond to the growth and propagation stages, respectively, and the six red curves correspond to $\lambda^*=2.412,~2.456,~2.536,~2.616,~2.696$ and $ 2.776.$}
\label{fig_maxwell}
\end{figure*}

\section{Characterisation of necking propagation}
To offer more insight into the propagation stage, we have shown in Fig.~\ref{fig_maxwell} a typical set of simulation results for the evolution of $\lambda_z$ versus $r$ as the outer radius is increased. It is seen that
 in the propagation stage (red curves), the $\lambda_z$ is a constant in the necked region, but seems to be slowly varying in the $r$-direction in the un-necked region. For easy reference, we refer to the necked region centered around the origin as the \lq\lq $-$ phase" and the other region with larger thickness as the \lq\lq $+$ phase". In this section, we give the two \lq\lq phases" an analytical description by assuming that $\lambda_z$ is a constant in both \lq\lq phases". Although the two \lq\lq phases" are connected by a thin transition region, we further assume that the transition region has zero thickness and simply refer to it as the interface.
We use superscripts \lq\lq +" and \lq\lq $-$" to signify evaluations at the \lq\lq $+$ phase" and \lq\lq $-$ phase" sides of the interface, respectively. Thus, for instance, $\lambda_1^+$ denotes the azimuthal stretch as the interface is approached from the \lq\lq $+$ phase" side.

The \lq\lq $-$ phase" is homogeneous. The solution is the same as in section 3, and it is uniquely determined by a single parameter which we take to be the stretch $\lambda_z^-$ in the $z$-direction. We then have $\lambda_1=\lambda_3=1/\sqrt{\lambda_z^-}$, and $\lambda_2= \lambda_z^-$.

For the \lq\lq $+$ phase", the solution is inhomogeneous although $\lambda_z^+$ is assumed to be a constant. In terms of cylindrical polar coordinates, the axisymmetric solution is given by
\be r=r(R), \;\;\;\;\theta=\Theta, \;\;\;\; z=\lambda_z^+ Z. \la{sep1} \en
The corresponding deformation gradient is given by $$F=r'(R) {\bm e}_r \otimes {\bm e}_r +(r/R) {\bm e}_\theta \otimes {\bm e}_\theta+ \lambda_z^+ {\bm e}_z \otimes {\bm e}_z, $$
so that $\lambda_1=r/R$, $\lambda_2=\lambda_z^+$, $\lambda_3=r'(R)$. To simplify notation, from now on we shall write $\lambda_z^+$ simply as $\lambda_z$.

By integrating the incompressibility condition det$\,F=1$, we obtain
\be r=\sqrt{\lambda_z^{-1} R^2+C}, \la{sep2} \en
where $C$ is a constant of integration. This constant may be determined by evaluating this expression at the edge of the plate $R=R_0$:
\be C=R_0^2 (\lambda^{*2} -\lambda_z^{-1}), \la{sep3} \en
where $\lambda^*$ denotes the azimuthal stretch at $R=R_0$ (so that $\lambda^*=r^*/R_0$ with $r^*$ denoting the deformed radius). Once the hydrostatic pressure has been eliminated with the use of $\sigma_2=0$, the principal Cauchy stresses $\sigma_1$ and $\sigma_3$ are given by
\be \sigma_1=\lambda_1 W_1-\lambda_2 W_2, \;\;\;\; \sigma_3=\lambda_3 W_3-\lambda_2 W_2. \la{sep4} \en
It then follows that
\be \sigma_3-\sigma_1= \lambda_1 w_{\lambda_1}, \la{sep5} \en
where $w_{\lambda_1}=d w/d\lambda_1$ and $w=w(\lambda_1)$ is the reduced strain energy function defined by
\be w(\lambda_1)=W(\lambda_1, \lambda_z, 1/(\lambda_z \lambda_1)). \la{sep6} \en
The only equilibrium equation that is not satisfied automatically is
\be \frac{d \sigma_3}{d r}+\frac{\sigma_3-\sigma_1}{r}=0. \la{sep7} \en
With use of the result \rr{sep5} and the relation $\lambda_1=r/R$, we may convert \rr{sep7} to the form (\cite{ho1979b})
\be \frac{d \sigma_3}{d \lambda_1}= \frac{w_{\lambda_1}}{\lambda_z \lambda_1^2-1}. \la{sep8} \en
On integrating this equation across the entire \lq\lq $+$ phase", we obtain
\be \sigma_3^*-\sigma_3^+=\int^{\lambda_1^*}_{\lambda_1^+} \frac{w_{\lambda_1}}{\lambda_z \lambda_1^2-1} d \lambda_1, \la{sep9} \en
where the superscript \lq\lq *"  signifies evaluation at the outer edge $r=r^*$.

The jump conditions that need to be imposed across the interface are
\be \sigma_3^+=\sigma_3^-, \;\;\;\; \lambda_1^+=\lambda_1^-, \;\;\;\; W^+-W^-= \sigma_3^- (\frac{\lambda_z^-}{\lambda_z}-1). \la{jump} \en
The first two correspond to traction and displacement continuity, whereas the third corresponds to stationarity of the total energy with respect to perturbations of the interface position in the undeformed configuration (see, e.g., \cite{gr1980}, \cite{fc1994}, or \cite{ff2004}).

With the use of \rr{jump}$_{1,2}$, equations \rr{sep9} and \rr{jump}$_{3}$ may be rewritten as
\be
\sigma_3^*-\sigma_3^-=\int^{\lambda_1^*}_{\lambda_1^-} \frac{w_{\lambda_1}}{\lambda_z \lambda_1^2-1} d \lambda_1,
\;\;\;\; W\left(\lambda_1^-, \lambda_z, \frac{1}{\lambda_1^- \lambda_z}\right)-W\left(\frac{1}{\sqrt{\lambda_z^-}}, \lambda_z^-, \sqrt{\lambda_z^-}\right)=\sigma_3^- \left(\frac{\lambda_z^-}{\lambda_z}-1\right).
\la{sep12} \en
These two equations can be used to find $\lambda_z$ and $\lambda_z^-$ for each specified $\lambda_1^*$ (noting that $\lambda_1^-=1/\sqrt{\lambda_z^-}$). The location of the interface, $r=r_s$ say, may be expressed in terms of $\lambda_z^-$ and $\lambda_z$ by solving the equation $\lambda^-=\lambda^+ $, that is,
\be
\frac{1}{\sqrt{\lambda_z^-}}=\frac{r_s}{\sqrt{\lambda_z r_s^2-\lambda_z C}}, \;\;\;\; \Longrightarrow \;\; \left(\frac{r_s}{R_0}\right)^2=\frac{\lambda_z \lambda_1^{*2}-1}{\lambda_z-\lambda_z^-}. \la{sep10} \en
Since we must have $0< r_s < R_0 \lambda_1^-$, the $\lambda_1^*$ must satisfy
\be 1/\sqrt{\lambda_z} < \lambda_1^* < 1/\sqrt{\lambda_z^-}. \la{sep13} \en
In Fig.~\ref{fig_maxwell}, we have shown the dependence of $\lambda_z$ ($=\lambda_z^+$) and $\lambda_z^-$ on $\lambda_1^*$, as predicted by \rr{sep12}, when
the strain-energy function is given by \rr{energy_2terms} with $m_1=1/2$, $m_2=4$ and $\mu_2/\mu_1=1/80$. On the same figure are shown six representative numerical simulation results (evaluated approximately at the center of the \lq\lq $+$ phase"). It is seen that there is excellent agreement for the values of $\lambda_z^-$ with relative error less than 0.57\%. The agreement for $\lambda_z^+$ is less satisfactory. We attribute this to the approximations that we have used, namely that the thickness of the transition region is neglected and the $\lambda_z$ in the  \lq\lq $+$ phase" is independent of $r$.
%

\begin{figure*}[ht]
\centering
\includegraphics[width=0.55\linewidth]{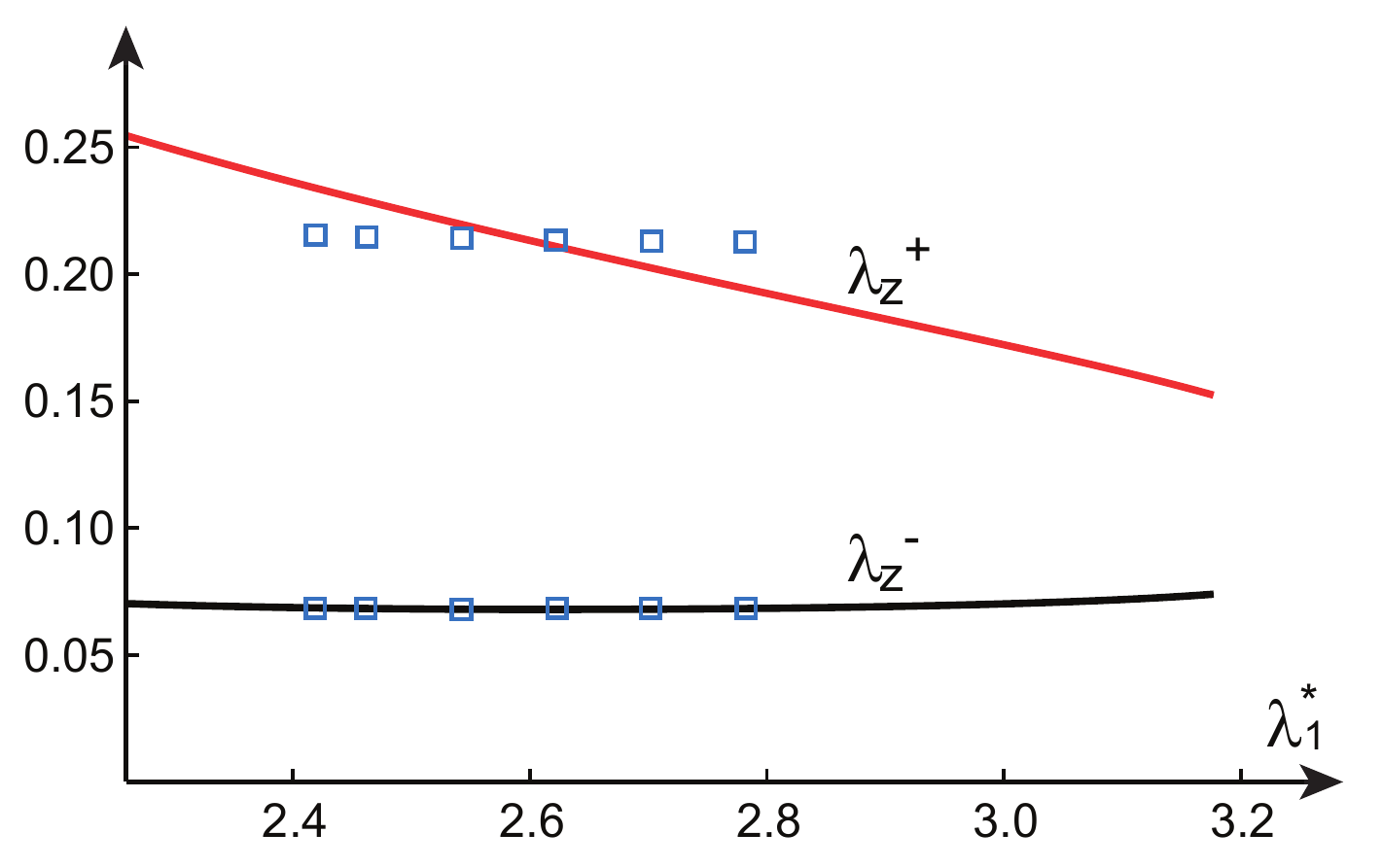}
\caption{Comparison of theoretical (solid lines) and numerical simulation results (squares) for $\lambda_z^+$ and $\lambda_z^-$ as functions of $\lambda_1^*$. The squares are corresponding to the six red lines in Figure \ref{fig_maxwell}.}
\label{fig_maxwell2}
\end{figure*}

\section{Conclusion}
For problems that have translational invariance in one spatial direction, e.g. the inflation of a homogeneous rubber tube or extension of a homogeneous plate in a state of plane strain, a rigorous analysis based on center manifold reduction can be conducted to establish the existence of a localised solution bifurcating from the trivial state \citep{ki1982, mi1991, hi2011}. A necessary condition for localisation has the simple interpretation that a static extensional mode can exist in the small wavenumber limit.
For the problem of localised bulging in an inflated rubber tube, this condition has been found to be equivalent to the Jacobian determinant of two force functions equal to zero. For problems defined over an infinite or semi-infinite domain that do not have translational invariance in the direction of localisation, such as the current axially symmetric necking problem, no center manifold reductions seem to have been attempted. We postulated that localised necking is still governed by the condition that an extensional bifurcation mode exists in the small wavenumber limit. Although we have not yet been able to ground this condition on any rigorous theory, we have been able to verify this postulate with the aid of Abaqus simulations. We further showed that the bifurcation condition for necking does not correspond to the Jacobian determinant equal to zero. The latter condition can always be factorised and the two factors give the conditions for the TK instability and limiting-point instability, respectively. It is also demonstrated, first numerically and then analytically, that necking evolution follows a growth/propagation process that is typical of similar localization problems. It is hoped that the current analysis will guide future studies of elastic localisations under multiple fields such as axisymmetric necking of dielectric membranes that are subjected to both electrical and mechanical loading.
%
%
%
%
%
%
%
%
%


\subsection*{Acknowledgements}
This work was supported by the National Natural Science Foundation of China (Grant No 12072224).


\bibliography{references}

\begin{thebibliography}{82}
\expandafter\ifx\csname natexlab\endcsname\relax\def\natexlab#1{#1}\fi
\providecommand{\url}[1]{\texttt{#1}}
\providecommand{\href}[2]{#2}
\providecommand{\path}[1]{#1}
\providecommand{\DOIprefix}{doi:}
\providecommand{\ArXivprefix}{arXiv:}
\providecommand{\URLprefix}{URL: }
\providecommand{\Pubmedprefix}{pmid:}
\providecommand{\doi}[1]{\href{http://dx.doi.org/#1}{\path{#1}}}
\providecommand{\Pubmed}[1]{\href{pmid:#1}{\path{#1}}}
\providecommand{\bibinfo}[2]{#2}
\ifx\xfnm\relax \def\xfnm[#1]{\unskip,\space#1}\fi
\bibitem[{{Abaqus}(2013)}]{ab2013}
\bibinfo{author}{{Abaqus}} (\bibinfo{year}{2013}).
\newblock {\it \bibinfo{title}{ABAQUS Analysis Users Manual, version 6.13}\/}.
\newblock \bibinfo{note}{Dassault Systems, Providence, RI, USA}.
\bibitem[{Antman(1972)}]{antman1972}
\bibinfo{author}{Antman, S.~S.} (\bibinfo{year}{1972}).
\newblock \bibinfo{title}{Qualitative theory of the ordinary differential
  equations of nonlinear elasticity}.
\newblock {\it \bibinfo{journal}{Mechanics Today}\/},  {\it
  \bibinfo{volume}{1}\/}, \bibinfo{pages}{58--101}.
\bibitem[{Antman(1973)}]{antman1973}
\bibinfo{author}{Antman, S.~S.} (\bibinfo{year}{1973}).
\newblock \bibinfo{title}{Nonuniqueness of equilbrium states for bars in
  tension}.
\newblock {\it \bibinfo{journal}{J. Math. Anal. Appl.}\/},  {\it
  \bibinfo{volume}{44}\/}, \bibinfo{pages}{333--349}.
\bibitem[{Antman \& Carbone(1977)}]{ac1977}
\bibinfo{author}{Antman, S.~S.}, \& \bibinfo{author}{Carbone, E.~R.}
  (\bibinfo{year}{1977}).
\newblock \bibinfo{title}{Shear and necking instability in nonlinear
  elasticity}.
\newblock {\it \bibinfo{journal}{J. Elast.}\/},  {\it \bibinfo{volume}{7}\/},
  \bibinfo{pages}{125--151}.
\bibitem[{Audoly \& Hutchinson(2016)}]{ah2016}
\bibinfo{author}{Audoly, B.}, \& \bibinfo{author}{Hutchinson, J.~W.}
  (\bibinfo{year}{2016}).
\newblock \bibinfo{title}{Analysis of necking based on a one-dimensional
  model}.
\newblock {\it \bibinfo{journal}{J. Mech. Phys. Solids}\/},  {\it
  \bibinfo{volume}{97}\/}, \bibinfo{pages}{68--91}.
\bibitem[{Baker \& Ericksen(1954)}]{be1954}
\bibinfo{author}{Baker, M.}, \& \bibinfo{author}{Ericksen, J.~L.}
  (\bibinfo{year}{1954}).
\newblock \bibinfo{title}{Inequalities restricting the form of stress
  deformation relations for isotropic elastic solids and reiner-rivlin fluids}.
\newblock {\it \bibinfo{journal}{J. Washington Acad. Sci.}\/},  {\it
  \bibinfo{volume}{44}\/}, \bibinfo{pages}{24--27}.
\bibitem[{Barenblatt(1964)}]{ba1964}
\bibinfo{author}{Barenblatt, G.~I.} (\bibinfo{year}{1964}).
\newblock \bibinfo{title}{On the neck propagation under tension of polymeric
  samples}.
\newblock {\it \bibinfo{journal}{Appl. Math. Mech.}\/},  {\it
  \bibinfo{volume}{28}\/}, \bibinfo{pages}{1048--1060}.
\bibitem[{Batra et~al.(2005)Batra, M\"uller \& Strehlow}]{bm2005}
\bibinfo{author}{Batra, R.~C.}, \bibinfo{author}{M\"uller, I.}, \&
  \bibinfo{author}{Strehlow, P.} (\bibinfo{year}{2005}).
\newblock \bibinfo{title}{Treloar’s biaxial tests and kearsley’s
  bifurcation in rubber sheets}.
\newblock {\it \bibinfo{journal}{Math. Mech. Solids}\/},  {\it
  \bibinfo{volume}{10}\/}, \bibinfo{pages}{705--713}.
\bibitem[{Bertoldi \& Gei(2011)}]{bg2011}
\bibinfo{author}{Bertoldi, K.}, \& \bibinfo{author}{Gei, M.}
  (\bibinfo{year}{2011}).
\newblock \bibinfo{title}{Instabilities in multilayered soft dielectrics}.
\newblock {\it \bibinfo{journal}{J. Mech. Phys. Solids}\/},  {\it
  \bibinfo{volume}{59}\/}, \bibinfo{pages}{18--42}.
\bibitem[{Bortot \& Shmuel(2018)}]{bs2018}
\bibinfo{author}{Bortot, E.}, \& \bibinfo{author}{Shmuel, G.}
  (\bibinfo{year}{2018}).
\newblock \bibinfo{title}{Prismatic bifurcations of soft dielectric tubes}.
\newblock {\it \bibinfo{journal}{Int. J. Eng. Sci.}\/},  {\it
  \bibinfo{volume}{124}\/}, \bibinfo{pages}{104--114}.
\bibitem[{Burke \& Nix(1979)}]{bn1979}
\bibinfo{author}{Burke, M.~A.}, \& \bibinfo{author}{Nix, W.~D.}
  (\bibinfo{year}{1979}).
\newblock \bibinfo{title}{A numerical study of necking in the plane tension
  test}.
\newblock {\it \bibinfo{journal}{Int. J. Solids Struct.}\/},  {\it
  \bibinfo{volume}{15}\/}, \bibinfo{pages}{379--393}.
\bibitem[{Carothers \& Hill(1932)}]{ch1932}
\bibinfo{author}{Carothers, W.~H.}, \& \bibinfo{author}{Hill, J.~W.}
  (\bibinfo{year}{1932}).
\newblock \bibinfo{title}{Studies of polymerization and ring formation xv.
  artificial fibers from synthetic linear condensation superpolymers}.
\newblock {\it \bibinfo{journal}{J. Am. Chem. Soc.}\/},  {\it
  \bibinfo{volume}{54}\/}, \bibinfo{pages}{1579--1587}.
\bibitem[{Chen et~al.(2021)Chen, Yang, Wang, Yang, Dayal \& Sharma}]{cy2021}
\bibinfo{author}{Chen, L.~L.}, \bibinfo{author}{Yang, X.},
  \bibinfo{author}{Wang, B.~L.}, \bibinfo{author}{Yang, S.~Y.},
  \bibinfo{author}{Dayal, K.}, \& \bibinfo{author}{Sharma, P.}
  (\bibinfo{year}{2021}).
\newblock \bibinfo{title}{The interplay between symmetry-breaking and
  symmetry-preserving bifurcations in soft dielectric films and the emergence
  of giant electro-actuation}.
\newblock {\it \bibinfo{journal}{Extr. Mech. Lett.}\/},  {\it
  \bibinfo{volume}{43}\/}, \bibinfo{pages}{101151}.
\bibitem[{Chen(1971)}]{chen1971}
\bibinfo{author}{Chen, W.~H.} (\bibinfo{year}{1971}).
\newblock \bibinfo{title}{Necking of a bar}.
\newblock {\it \bibinfo{journal}{Int. J. Solids Struct.}\/},  {\it
  \bibinfo{volume}{7}\/}, \bibinfo{pages}{685--717}.
\bibitem[{Chen(1987)}]{chen1987}
\bibinfo{author}{Chen, Y.~C.} (\bibinfo{year}{1987}).
\newblock \bibinfo{title}{Stability of homogeneous deformations of an
  incompressible elastic body under dead-load surface tractions}.
\newblock {\it \bibinfo{journal}{J. Elast.}\/},  {\it \bibinfo{volume}{17}\/},
  \bibinfo{pages}{223--248}.
\bibitem[{Coleman(1983)}]{coleman1988}
\bibinfo{author}{Coleman, B.~D.} (\bibinfo{year}{1983}).
\newblock \bibinfo{title}{Necking and drawing in polymeric fibers under
  tension}.
\newblock {\it \bibinfo{journal}{Arch. Ratl Mech. Anal.}\/},  {\it
  \bibinfo{volume}{83}\/}, \bibinfo{pages}{115--137}.
\bibitem[{Coleman \& Newman(1988)}]{cn1988}
\bibinfo{author}{Coleman, B.~D.}, \& \bibinfo{author}{Newman, D.~C.}
  (\bibinfo{year}{1988}).
\newblock \bibinfo{title}{On the rheology of cold drawing. i. elastic
  materials}.
\newblock {\it \bibinfo{journal}{J. Polymer Sci.}\/},  {\it
  \bibinfo{volume}{26}\/}, \bibinfo{pages}{1801--1822}.
\bibitem[{Crissman \& Zapas(1974)}]{cz1974}
\bibinfo{author}{Crissman, J.~M.}, \& \bibinfo{author}{Zapas, I.~J.}
  (\bibinfo{year}{1974}).
\newblock \bibinfo{title}{Creep failure and fracture of polyethylene in
  uniaxial extension}.
\newblock {\it \bibinfo{journal}{Polymer Eng. Sci.}\/},  {\it
  \bibinfo{volume}{19}\/}, \bibinfo{pages}{99--103}.
\bibitem[{Dai \& Bi(2006)}]{db2006}
\bibinfo{author}{Dai, H.-H.}, \& \bibinfo{author}{Bi, Q.~S.}
  (\bibinfo{year}{2006}).
\newblock \bibinfo{title}{On constructing the unique solution for the necking
  in a hyper-elastic rod}.
\newblock {\it \bibinfo{journal}{J. Elast.}\/},  {\it \bibinfo{volume}{82}\/},
  \bibinfo{pages}{215--241}.
\bibitem[{Dai et~al.(2008)Dai, Hao \& Chen}]{dh2008}
\bibinfo{author}{Dai, H.-H.}, \bibinfo{author}{Hao, Y.~H.}, \&
  \bibinfo{author}{Chen, Z.} (\bibinfo{year}{2008}).
\newblock \bibinfo{title}{On constructing the analytical solutions for
  localizations in a slender cylinder composed of an incompressible
  hyperelastic material}.
\newblock {\it \bibinfo{journal}{Int. J. Solids Struct.}\/},  {\it
  \bibinfo{volume}{45}\/}, \bibinfo{pages}{2613--2628}.
\bibitem[{Dai \& Peng(2012)}]{dp2012}
\bibinfo{author}{Dai, H.-H.}, \& \bibinfo{author}{Peng, X.~C.}
  (\bibinfo{year}{2012}).
\newblock \bibinfo{title}{Elliptic-spline solutions for large localizations in
  a circular blatz-ko cylinder due to geometric softening}.
\newblock {\it \bibinfo{journal}{SIAM J. Appl. Math.}\/},  {\it
  \bibinfo{volume}{72}\/}, \bibinfo{pages}{181--200}.
\bibitem[{Dorfmann \& Ogden(2014)}]{do2014}
\bibinfo{author}{Dorfmann, L.}, \& \bibinfo{author}{Ogden, R.~W.}
  (\bibinfo{year}{2014}).
\newblock \bibinfo{title}{Instabilities of an electroelastic plate}.
\newblock {\it \bibinfo{journal}{Int. J. Eng. Sci.}\/},  {\it
  \bibinfo{volume}{77}\/}, \bibinfo{pages}{79--101}.
\bibitem[{Dorfmann \& Ogden(2019)}]{do2019}
\bibinfo{author}{Dorfmann, L.}, \& \bibinfo{author}{Ogden, R.~W.}
  (\bibinfo{year}{2019}).
\newblock \bibinfo{title}{Instabilities of soft dielectrics}.
\newblock {\it \bibinfo{journal}{Phil. Trans. R. Soc. A}\/},  {\it
  \bibinfo{volume}{377}\/}, \bibinfo{pages}{20180077}.
\bibitem[{Dowaikh \& Ogden(1990)}]{do1990}
\bibinfo{author}{Dowaikh, M.~A.}, \& \bibinfo{author}{Ogden, R.~W.}
  (\bibinfo{year}{1990}).
\newblock \bibinfo{title}{On surface waves and deformations in a pre-stressed
  incompressible elastic solid}.
\newblock {\it \bibinfo{journal}{IMA J. Appl. Math.}\/},  {\it
  \bibinfo{volume}{44}\/}, \bibinfo{pages}{261--284}.
\bibitem[{Ericksen(1975)}]{er1975}
\bibinfo{author}{Ericksen, J.~L.} (\bibinfo{year}{1975}).
\newblock \bibinfo{title}{Equilibrium of bars}.
\newblock {\it \bibinfo{journal}{J. Elast.}\/},  {\it \bibinfo{volume}{5}\/},
  \bibinfo{pages}{191--201}.
\bibitem[{Freidin \& Chiskis(1994)}]{fc1994}
\bibinfo{author}{Freidin, A.~B.}, \& \bibinfo{author}{Chiskis, A.~M.}
  (\bibinfo{year}{1994}).
\newblock \bibinfo{title}{Phase transition zones in nonlinear elastic isotropic
  materials}.
\newblock {\it \bibinfo{journal}{Izv. AN USSR Mekh. Tverdogo Tela}\/},  {\it
  \bibinfo{volume}{29}\/}, \bibinfo{pages}{91--109}.
\bibitem[{Fu(2001)}]{fu2001}
\bibinfo{author}{Fu, Y.~B.} (\bibinfo{year}{2001}).
\newblock {\it \bibinfo{title}{Nonlinear stability analysis. In Nonlinear
  elasticity: theory and applications (eds YB Fu, RW Ogden)}\/}.
\newblock \bibinfo{publisher}{Cambridge University Press, Cambridge}.
\bibitem[{Fu et~al.(2018{\natexlab{a}})Fu, Dorfmann \& Xie}]{fdx2018}
\bibinfo{author}{Fu, Y.~B.}, \bibinfo{author}{Dorfmann, L.}, \&
  \bibinfo{author}{Xie, Y.~X.} (\bibinfo{year}{2018}{\natexlab{a}}).
\newblock \bibinfo{title}{Localized necking of a dielectric membrane}.
\newblock {\it \bibinfo{journal}{Extr. Mech. Lett.}\/},  {\it
  \bibinfo{volume}{21}\/}, \bibinfo{pages}{44--48}.
\bibitem[{Fu \& Freidin(2004)}]{ff2004}
\bibinfo{author}{Fu, Y.~B.}, \& \bibinfo{author}{Freidin, A.~B.}
  (\bibinfo{year}{2004}).
\newblock \bibinfo{title}{Characterization and stability of two-phase
  piecewise-homogeneous deformations}.
\newblock {\it \bibinfo{journal}{Proc. R. Soc. A}\/},  {\it
  \bibinfo{volume}{460}\/}, \bibinfo{pages}{3065--3094}.
\bibitem[{Fu et~al.(2021)Fu, Jin \& Goriely}]{fjg2021}
\bibinfo{author}{Fu, Y.~B.}, \bibinfo{author}{Jin, L.}, \&
  \bibinfo{author}{Goriely, A.} (\bibinfo{year}{2021}).
\newblock \bibinfo{title}{Necking, beading, and bulging in soft elastic
  cylinders.}
\newblock {\it \bibinfo{journal}{J. Mech. Phys. Solids}\/},  {\it
  \bibinfo{volume}{147}\/}, \bibinfo{pages}{104250}.
\bibitem[{Fu et~al.(2016)Fu, Liu \& Francisco}]{fl2016}
\bibinfo{author}{Fu, Y.~B.}, \bibinfo{author}{Liu, J.~L.}, \&
  \bibinfo{author}{Francisco, G.~S.} (\bibinfo{year}{2016}).
\newblock \bibinfo{title}{Localized bulging in an inflated cylindrical tube of
  arbitrary thickness -- the effect of bending stiffness}.
\newblock {\it \bibinfo{journal}{J. Mech. Phys. Solids}\/},  {\it
  \bibinfo{volume}{90}\/}, \bibinfo{pages}{45--60}.
\bibitem[{Fu \& Ogden(1999)}]{fo1999}
\bibinfo{author}{Fu, Y.~B.}, \& \bibinfo{author}{Ogden, R.~W.}
  (\bibinfo{year}{1999}).
\newblock \bibinfo{title}{Nonlinear stability analysis of pre-stressed elastic
  bodies}.
\newblock {\it \bibinfo{journal}{Cont. Mech. Thermodyn.}\/},  {\it
  \bibinfo{volume}{11}\/}, \bibinfo{pages}{141--172}.
\bibitem[{Fu et~al.(2008)Fu, Pearce \& Liu}]{fpl2008}
\bibinfo{author}{Fu, Y.~B.}, \bibinfo{author}{Pearce, S.~P.}, \&
  \bibinfo{author}{Liu, K.-K.} (\bibinfo{year}{2008}).
\newblock \bibinfo{title}{Post-bifurcation analysis of a thin-walled
  hyperelastic tube under inflation}.
\newblock {\it \bibinfo{journal}{Int. J. Non-linear Mech.}\/},  {\it
  \bibinfo{volume}{43}\/}, \bibinfo{pages}{697--706}.
\bibitem[{Fu \& Rogerson(1994)}]{fr1994}
\bibinfo{author}{Fu, Y.~B.}, \& \bibinfo{author}{Rogerson, G.~A.}
  (\bibinfo{year}{1994}).
\newblock \bibinfo{title}{A nonlinear analysis of instability of a pre-stressed
  incompressible elastic platet}.
\newblock {\it \bibinfo{journal}{Proc. R. Soc. Lond. A}\/},  {\it
  \bibinfo{volume}{446}\/}, \bibinfo{pages}{233--254}.
\bibitem[{Fu \& Xie(2010)}]{fx2010}
\bibinfo{author}{Fu, Y.~B.}, \& \bibinfo{author}{Xie, Y.~X.}
  (\bibinfo{year}{2010}).
\newblock \bibinfo{title}{Stability of localized bulging in inflated membrane
  tubes under volume control}.
\newblock {\it \bibinfo{journal}{Int. J. Eng. Sci.}\/},  {\it
  \bibinfo{volume}{48}\/}, \bibinfo{pages}{1242--1252}.
\bibitem[{Fu et~al.(2018{\natexlab{b}})Fu, Xie \& Dorfmann}]{fxd2018}
\bibinfo{author}{Fu, Y.~B.}, \bibinfo{author}{Xie, Y.~X.}, \&
  \bibinfo{author}{Dorfmann, L.} (\bibinfo{year}{2018}{\natexlab{b}}).
\newblock \bibinfo{title}{A reduced model for electrodes-coated dielectric
  plates}.
\newblock {\it \bibinfo{journal}{Int. J. Non-linear Mech.}\/},  {\it
  \bibinfo{volume}{106}\/}, \bibinfo{pages}{60--69}.
\bibitem[{Giudici \& Biggins(2020)}]{giudici2020}
\bibinfo{author}{Giudici, A.}, \& \bibinfo{author}{Biggins, J.~S.}
  (\bibinfo{year}{2020}).
\newblock \bibinfo{title}{Ballooning, bulging and necking: an exact solution
  for longitudinal phase separation in elastic systems near a critical point}.
\newblock {\it \bibinfo{journal}{Phys. Rev. E}\/},  {\it
  \bibinfo{volume}{102}\/}, \bibinfo{pages}{033007}.
\bibitem[{Grinfeld(1980)}]{gr1980}
\bibinfo{author}{Grinfeld, M.~A.} (\bibinfo{year}{1980}).
\newblock \bibinfo{title}{On conditions of thermodynamic equilibrium of the
  phases of a nonlinear elastic material}.
\newblock {\it \bibinfo{journal}{Dokl. Akad. Nauk SSSR}\/},  {\it
  \bibinfo{volume}{251}\/}, \bibinfo{pages}{824--827}.
\bibitem[{Haragus \& Iooss(2011)}]{hi2011}
\bibinfo{author}{Haragus, M.}, \& \bibinfo{author}{Iooss, G.}
  (\bibinfo{year}{2011}).
\newblock {\it \bibinfo{title}{Local bifurcations, center manifolds, and normal
  forms in infinite-dimensional dynamical systems}\/}.
\newblock \bibinfo{publisher}{Springer, London}.
\bibitem[{Haughton \& Ogden(1979)}]{ho1979b}
\bibinfo{author}{Haughton, D.~M.}, \& \bibinfo{author}{Ogden, R.~W.}
  (\bibinfo{year}{1979}).
\newblock \bibinfo{title}{Bifurcation of inflated circular cylinders of elastic
  material under axial loading ii. exact theory for thick-walled tubes}.
\newblock {\it \bibinfo{journal}{J. Mech. Phys. Solids}\/},  {\it
  \bibinfo{volume}{27}\/}, \bibinfo{pages}{489--512}.
\bibitem[{Hill \& Hutchinson(1975)}]{hh1975}
\bibinfo{author}{Hill, R.}, \& \bibinfo{author}{Hutchinson, J.~W.}
  (\bibinfo{year}{1975}).
\newblock \bibinfo{title}{Bifurcation phenomena in the plane tension test}.
\newblock {\it \bibinfo{journal}{J. Mech. Phys. Solids}\/},  {\it
  \bibinfo{volume}{23}\/}, \bibinfo{pages}{239–264}.
\bibitem[{Hutchinson \& Miles(1974)}]{hm1974}
\bibinfo{author}{Hutchinson, J.~W.}, \& \bibinfo{author}{Miles, J.~P.}
  (\bibinfo{year}{1974}).
\newblock \bibinfo{title}{Bifurcation analysis of the onset of necking in an
  elastic/plastic cylinder under uniaxial tension}.
\newblock {\it \bibinfo{journal}{J. Mech. Phys. Solids}\/},  {\it
  \bibinfo{volume}{22}\/}, \bibinfo{pages}{61--71}.
\bibitem[{Kearsley(1986)}]{ke1986}
\bibinfo{author}{Kearsley, E.~A.} (\bibinfo{year}{1986}).
\newblock \bibinfo{title}{Asymmetric stretching of a symmetrically loaded
  elastic sheet}.
\newblock {\it \bibinfo{journal}{Int. J. Solids Struct.}\/},  {\it
  \bibinfo{volume}{22}\/}, \bibinfo{pages}{111--119}.
\bibitem[{Kirchg\"assner(1982)}]{ki1982}
\bibinfo{author}{Kirchg\"assner, K.} (\bibinfo{year}{1982}).
\newblock \bibinfo{title}{Wave solutions of reversible systems and
  applications}.
\newblock {\it \bibinfo{journal}{J. Diff. Eqns.}\/},  {\it
  \bibinfo{volume}{45}\/}, \bibinfo{pages}{113--127}.
\bibitem[{Kyriakides \& Chang(1991)}]{kc1991}
\bibinfo{author}{Kyriakides, S.}, \& \bibinfo{author}{Chang, Y.-C.}
  (\bibinfo{year}{1991}).
\newblock \bibinfo{title}{The initiation and propagation of a localized
  instability in an inflated elastic tube}.
\newblock {\it \bibinfo{journal}{Int. J. Solids Struct.}\/},  {\it
  \bibinfo{volume}{27}\/}, \bibinfo{pages}{1085--1111}.
\bibitem[{Lestringant \& Audoly(2020)}]{la2020}
\bibinfo{author}{Lestringant, C.}, \& \bibinfo{author}{Audoly, B.}
  (\bibinfo{year}{2020}).
\newblock \bibinfo{title}{A one-dimensional model for elasto-capillary
  necking}.
\newblock {\it \bibinfo{journal}{Proc. Roy. Soc. A}\/},  {\it
  \bibinfo{volume}{476}\/}, \bibinfo{pages}{20200337}.
\bibitem[{Lu et~al.(2020)Lu, Ma \& Wang}]{lm2020}
\bibinfo{author}{Lu, T.~Q.}, \bibinfo{author}{Ma, C.}, \&
  \bibinfo{author}{Wang, T.~J.} (\bibinfo{year}{2020}).
\newblock \bibinfo{title}{Mechanics of dielectric elastomer structures: A
  review}.
\newblock {\it \bibinfo{journal}{Extr. Mech. Lett.}\/},  {\it
  \bibinfo{volume}{38}\/}, \bibinfo{pages}{100752}.
\bibitem[{MacSithigh(1986)}]{mac1986}
\bibinfo{author}{MacSithigh, G.~P.} (\bibinfo{year}{1986}).
\newblock \bibinfo{title}{Energy-minimal finite deformations of a symmetrically
  loaded elastic sheet}.
\newblock {\it \bibinfo{journal}{Q. J. Mech. Appl. Maths}\/},  {\it
  \bibinfo{volume}{39}\/}, \bibinfo{pages}{111--123}.
\bibitem[{MacSithigh \& Chen(1992)}]{mc1992}
\bibinfo{author}{MacSithigh, G.~P.}, \& \bibinfo{author}{Chen, Y.~C.}
  (\bibinfo{year}{1992}).
\newblock \bibinfo{title}{Bifurcation and stability of an incompressible
  elastic body under homogeneous dead loads with symmetry. part i: General
  isotropic materials}.
\newblock {\it \bibinfo{journal}{Q. J. Mech. Appl. Maths}\/},  {\it
  \bibinfo{volume}{45}\/}, \bibinfo{pages}{277--291}.
\bibitem[{Mielke(1991)}]{mi1991}
\bibinfo{author}{Mielke, A.} (\bibinfo{year}{1991}).
\newblock {\it \bibinfo{title}{Hamiltonian and Lagrangian Flows on Center
  Manifolds, with Applications to Elliptic Variational Problems}\/}.
\newblock \bibinfo{publisher}{Springer-Verlag (Lecture Notes in Mathematics
  vol. 1489), Berlin}.
\bibitem[{Mora et~al.(2010)Mora, Phou, Fromental, Pismen \& Pomeau}]{mp2010}
\bibinfo{author}{Mora, S.}, \bibinfo{author}{Phou, T.},
  \bibinfo{author}{Fromental, J.-M.}, \bibinfo{author}{Pismen, L.~M.}, \&
  \bibinfo{author}{Pomeau, Y.} (\bibinfo{year}{2010}).
\newblock \bibinfo{title}{Capillarity driven instability of a soft solid}.
\newblock {\it \bibinfo{journal}{Phys. Rev. Lett}\/},  {\it
  \bibinfo{volume}{105}\/}, \bibinfo{pages}{214301}.
\bibitem[{Na et~al.(2006)Na, Tanaka, Kawauchi, Furukawa, Sumiyoshi, Gong \&
  Osada}]{ntk2006}
\bibinfo{author}{Na, Y.~H.}, \bibinfo{author}{Tanaka, Y.},
  \bibinfo{author}{Kawauchi, Y.}, \bibinfo{author}{Furukawa, H.},
  \bibinfo{author}{Sumiyoshi, T.}, \bibinfo{author}{Gong, J.~P.}, \&
  \bibinfo{author}{Osada, Y.} (\bibinfo{year}{2006}).
\newblock \bibinfo{title}{Necking phenomenon of double-network gels}.
\newblock {\it \bibinfo{journal}{Macromolecules}\/},  {\it
  \bibinfo{volume}{39}\/}, \bibinfo{pages}{4641--4645}.
\bibitem[{Needleman(1972)}]{ne1972}
\bibinfo{author}{Needleman, A.} (\bibinfo{year}{1972}).
\newblock \bibinfo{title}{A numerical study of necking in circular cylindrical
  bar}.
\newblock {\it \bibinfo{journal}{J. Mech. Phys. Solids}\/},  {\it
  \bibinfo{volume}{20}\/}, \bibinfo{pages}{111--127}.
\bibitem[{Norris(2008)}]{no2008}
\bibinfo{author}{Norris, A.~N.} (\bibinfo{year}{2008}).
\newblock \bibinfo{title}{Comment on method to analyze electromechanical
  stability of dielectric elastomers, appl. phys. lett. 91 (2007) 061921}.
\newblock {\it \bibinfo{journal}{Appl. Phys. Lett.}\/},  {\it
  \bibinfo{volume}{92}\/}, \bibinfo{pages}{026101}.
\bibitem[{Norris et~al.(1978)Norris, Moran, Scudder \& Quinones}]{nm1978}
\bibinfo{author}{Norris, D.~M.}, \bibinfo{author}{Moran, B.},
  \bibinfo{author}{Scudder, J.~K.}, \& \bibinfo{author}{Quinones, D.~F.}
  (\bibinfo{year}{1978}).
\newblock \bibinfo{title}{A computer simulation of the tension test}.
\newblock {\it \bibinfo{journal}{J. Mech. Phys. Solids}\/},  {\it
  \bibinfo{volume}{26}\/}, \bibinfo{pages}{1--19}.
\bibitem[{Ogden(1987)}]{og1987}
\bibinfo{author}{Ogden} (\bibinfo{year}{1987}).
\newblock \bibinfo{title}{On the stability of asymmetric deformations of a
  symmetrically-tensioned elastic sheet}.
\newblock {\it \bibinfo{journal}{Int. J. Eng. Sci.}\/},  {\it
  \bibinfo{volume}{25}\/}, \bibinfo{pages}{1305--1314}.
\bibitem[{Ogden(1984)}]{og1984}
\bibinfo{author}{Ogden, R.~W.} (\bibinfo{year}{1984}).
\newblock {\it \bibinfo{title}{Non-linear Elastic Deformations}\/}.
\newblock \bibinfo{publisher}{Ellis Horwood, New York}.
\bibitem[{Ogden(1985)}]{Og1985}
\bibinfo{author}{Ogden, R.~W.} (\bibinfo{year}{1985}).
\newblock \bibinfo{title}{Local and global bifurcation phenomena in
  plane-strain finite elasticity}.
\newblock {\it \bibinfo{journal}{Int. J. Solids Struct.}\/},  {\it
  \bibinfo{volume}{21}\/}, \bibinfo{pages}{121--132}.
\bibitem[{Owen(1987)}]{owen1987}
\bibinfo{author}{Owen, N.} (\bibinfo{year}{1987}).
\newblock \bibinfo{title}{Existence and stability of necking deformations for
  nonlinearly elastic rods}.
\newblock {\it \bibinfo{journal}{Arch. Ratl. Mech. Anal.}\/},  {\it
  \bibinfo{volume}{98}\/}, \bibinfo{pages}{357--383}.
\bibitem[{Puglisi \& Zurlo(2012)}]{pz2012}
\bibinfo{author}{Puglisi, G.}, \& \bibinfo{author}{Zurlo, G.}
  (\bibinfo{year}{2012}).
\newblock \bibinfo{title}{Catastrophic thinning of dielectric elastomers}.
\newblock {\it \bibinfo{journal}{J. Electrostat}\/},  {\it
  \bibinfo{volume}{70}\/}, \bibinfo{pages}{312–316}.
\bibitem[{Riks(1979)}]{riks1979}
\bibinfo{author}{Riks, E.} (\bibinfo{year}{1979}).
\newblock \bibinfo{title}{An incremental approach to the solution of snapping
  and buckling problems}.
\newblock {\it \bibinfo{journal}{Int. J. Solids Struct.}\/},  {\it
  \bibinfo{volume}{15}\/}, \bibinfo{pages}{529--551}.
\bibitem[{Scherzinger \& Triantafyllidis(1998)}]{st1998}
\bibinfo{author}{Scherzinger, W.}, \& \bibinfo{author}{Triantafyllidis, N.}
  (\bibinfo{year}{1998}).
\newblock \bibinfo{title}{Asymptotic analysis of stability for prismatic solids
  under axial loads}.
\newblock {\it \bibinfo{journal}{J. Mech. Phys. Solids}\/},  {\it
  \bibinfo{volume}{46}\/}, \bibinfo{pages}{955--1007}.
\bibitem[{Silling(1988)}]{si1988}
\bibinfo{author}{Silling, S.~A.} (\bibinfo{year}{1988}).
\newblock \bibinfo{title}{Two-dimensional effects in the necking of elastic
  bars}.
\newblock {\it \bibinfo{journal}{J. Appl. Mech.}\/},  {\it
  \bibinfo{volume}{55}\/}, \bibinfo{pages}{530--535}.
\bibitem[{Steigmann(2007)}]{st2007}
\bibinfo{author}{Steigmann, D.~J.} (\bibinfo{year}{2007}).
\newblock \bibinfo{title}{A simple model of the treloar-kearsley instability}.
\newblock {\it \bibinfo{journal}{Math. Mech. Solids}\/},  {\it
  \bibinfo{volume}{12}\/}, \bibinfo{pages}{611--622}.
\bibitem[{Su et~al.(2018)Su, Broderick, Chen \& Destrade}]{sb2018}
\bibinfo{author}{Su, Y.~P.}, \bibinfo{author}{Broderick, H.~C.},
  \bibinfo{author}{Chen, W.~Q.}, \& \bibinfo{author}{Destrade, M.}
  (\bibinfo{year}{2018}).
\newblock \bibinfo{title}{Wrinkles in soft dielectric plates}.
\newblock {\it \bibinfo{journal}{J. Mech. Phys. Solids}\/},  {\it
  \bibinfo{volume}{119}\/}, \bibinfo{pages}{298–318}.
\bibitem[{Treloar(1949)}]{tr1949}
\bibinfo{author}{Treloar, L. R.~G.} (\bibinfo{year}{1949}).
\newblock {\it \bibinfo{title}{The Physics of Rubber Elasticity}\/}.
\newblock \bibinfo{publisher}{Clarendon Press, Oxford}.
\bibitem[{Triantafyllidis \& Aifantis(1986)}]{ta1986}
\bibinfo{author}{Triantafyllidis, N.}, \& \bibinfo{author}{Aifantis, E.~C.}
  (\bibinfo{year}{1986}).
\newblock \bibinfo{title}{A gradient approach to localization of deformation.
  i. hyperelastic materials}.
\newblock {\it \bibinfo{journal}{J. Elast.}\/},  {\it \bibinfo{volume}{16}\/},
  \bibinfo{pages}{225--237}.
\bibitem[{Triantafyllidis et~al.(2007)Triantafyllidis, Scherzinger \&
  Huang}]{tr2007}
\bibinfo{author}{Triantafyllidis, N.}, \bibinfo{author}{Scherzinger, W.~M.}, \&
  \bibinfo{author}{Huang, H.-J.} (\bibinfo{year}{2007}).
\newblock \bibinfo{title}{Post-bifurcation equilibria in the plane-strain test
  of a hyperelastic rectangular block}.
\newblock {\it \bibinfo{journal}{Int. J. Solids Struct.}\/},  {\it
  \bibinfo{volume}{44}\/}, \bibinfo{pages}{3700--3719}.
\bibitem[{Wang \& Fu(2021)}]{wf2021}
\bibinfo{author}{Wang, M.}, \& \bibinfo{author}{Fu, Y.~B.}
  (\bibinfo{year}{2021}).
\newblock \bibinfo{title}{Necking of a hyperelastic solid cylinder under axial
  stretching: Evaluation of the infinite-length approximation}.
\newblock {\it \bibinfo{journal}{Int. J. Eng. Sci.}\/},  {\it
  \bibinfo{volume}{159}\/}, \bibinfo{pages}{103432}.
\bibitem[{Wang et~al.(2019)Wang, Guo, Zhou, Li \& Fu}]{wg2019}
\bibinfo{author}{Wang, S.~B.}, \bibinfo{author}{Guo, G.~M.},
  \bibinfo{author}{Zhou, L.}, \bibinfo{author}{Li, L.~A.}, \&
  \bibinfo{author}{Fu, Y.~B.} (\bibinfo{year}{2019}).
\newblock \bibinfo{title}{An experimental study of localized bulging in
  inflated cylindrical tubes guided by newly emerged analytical results}.
\newblock {\it \bibinfo{journal}{J. Mech. Phys. Solids}\/},  {\it
  \bibinfo{volume}{124}\/}, \bibinfo{pages}{536--554}.
\bibitem[{Ward \& Sweeney(1982)}]{ward2013}
\bibinfo{author}{Ward, I.~M.}, \& \bibinfo{author}{Sweeney, J.}
  (\bibinfo{year}{1982}).
\newblock {\it \bibinfo{title}{Mechanical Properties of Solid Polymers (3rd
  Ed.)}\/}.
\newblock \bibinfo{publisher}{Wiley, New York}.
\bibitem[{Wesolowski(1962)}]{wes1962}
\bibinfo{author}{Wesolowski, Z.} (\bibinfo{year}{1962}).
\newblock \bibinfo{title}{Stability in some cases of tension in the light of
  the theory of finite strain}.
\newblock {\it \bibinfo{journal}{Arch. Mech. Stos.}\/},  {\it
  \bibinfo{volume}{14}\/}, \bibinfo{pages}{875--900}.
\bibitem[{Whitney \& Andrews(1967)}]{wa1967}
\bibinfo{author}{Whitney, W.}, \& \bibinfo{author}{Andrews, R.~D.}
  (\bibinfo{year}{1967}).
\newblock \bibinfo{title}{Yielding of glassy polymers: volume effects}.
\newblock {\it \bibinfo{journal}{J. Polym. Sci. Part C}\/},  {\it
  \bibinfo{volume}{16}\/}, \bibinfo{pages}{2981--2990}.
\bibitem[{Xia et~al.(2021)Xia, Su \& Chen}]{xsc2021}
\bibinfo{author}{Xia, G.~Z.}, \bibinfo{author}{Su, Y.~P.}, \&
  \bibinfo{author}{Chen, W.~Q.} (\bibinfo{year}{2021}).
\newblock \bibinfo{title}{Instability of compressible soft electroactive
  plates}.
\newblock {\it \bibinfo{journal}{Int. J. Eng. Sci.}\/},  {\it
  \bibinfo{volume}{162}\/}, \bibinfo{pages}{103474}.
\bibitem[{Xuan \& Biggins(2017)}]{xuan2017}
\bibinfo{author}{Xuan, C.}, \& \bibinfo{author}{Biggins, J.}
  (\bibinfo{year}{2017}).
\newblock \bibinfo{title}{Plateau-rayleigh instability in solids is a simple
  phase separation}.
\newblock {\it \bibinfo{journal}{Phys. Rev. E}\/},  {\it
  \bibinfo{volume}{95}\/}, \bibinfo{pages}{053106}.
\bibitem[{Yang et~al.(2017)Yang, Zhao \& Sharma}]{yzs2017}
\bibinfo{author}{Yang, S.~Y.}, \bibinfo{author}{Zhao, X.~H.}, \&
  \bibinfo{author}{Sharma, P.} (\bibinfo{year}{2017}).
\newblock \bibinfo{title}{Revisiting the instability and bifurcation behavior
  of soft dielectrics}.
\newblock {\it \bibinfo{journal}{J. Appl. Mech.}\/},  {\it
  \bibinfo{volume}{84}\/}, \bibinfo{pages}{031008}.
\bibitem[{Ye et~al.(2019)Ye, Liu \& Fu}]{ylf2019}
\bibinfo{author}{Ye, Y.}, \bibinfo{author}{Liu, Y.}, \& \bibinfo{author}{Fu,
  Y.~B.} (\bibinfo{year}{2019}).
\newblock \bibinfo{title}{Weakly nonlinear analysis of localized bulging of an
  inflated hyperelastic tube of arbitrary wall thickness}.
\newblock {\it \bibinfo{journal}{J. Mech. Phys. Solids}\/},  {\it
  \bibinfo{volume}{135}\/}, \bibinfo{pages}{103804}.
\bibitem[{Zapas \& Crissman(1974)}]{zc1974}
\bibinfo{author}{Zapas, I.~J.}, \& \bibinfo{author}{Crissman, J.~M.}
  (\bibinfo{year}{1974}).
\newblock \bibinfo{title}{An instability leading to failure of polyethylene in
  uniaxial creep}.
\newblock {\it \bibinfo{journal}{Polym. Eng. Sci.}\/},  {\it
  \bibinfo{volume}{19}\/}, \bibinfo{pages}{104--107}.
\bibitem[{Zhang et~al.(2018)Zhang, Bai, Lei \& R}]{zb2018}
\bibinfo{author}{Zhang, R.}, \bibinfo{author}{Bai, P.~X.},
  \bibinfo{author}{Lei, D.}, \& \bibinfo{author}{R, X.} (\bibinfo{year}{2018}).
\newblock \bibinfo{title}{Aging-dependent strain localization in amorphous
  glassy polymers: From necking to shear banding}.
\newblock {\it \bibinfo{journal}{Int. J. Solids Struct.}\/},  {\it
  \bibinfo{volume}{146}\/}, \bibinfo{pages}{203--213}.
\bibitem[{Zhao(2012)}]{zhao2012}
\bibinfo{author}{Zhao, X.~H.} (\bibinfo{year}{2012}).
\newblock \bibinfo{title}{A theory for large deformation and damage of
  interpenetrating polymer networks}.
\newblock {\it \bibinfo{journal}{J. Mech. Phys. Solids}\/},  {\it
  \bibinfo{volume}{60}\/}, \bibinfo{pages}{319--332}.
\bibitem[{Zhao \& Suo(2007)}]{zs2007}
\bibinfo{author}{Zhao, X.~H.}, \& \bibinfo{author}{Suo, Z.}
  (\bibinfo{year}{2007}).
\newblock \bibinfo{title}{Method to analyze electromechanical stability of
  dielectric elastomers}.
\newblock {\it \bibinfo{journal}{Appl. Phys. Lett.}\/},  {\it
  \bibinfo{volume}{91}\/}, \bibinfo{pages}{061921}.
\bibitem[{Zurlo et~al.(2017)Zurlo, Destrade, DeTommasi \& Puglisi}]{zd2017}
\bibinfo{author}{Zurlo, G.}, \bibinfo{author}{Destrade, M.},
  \bibinfo{author}{DeTommasi, D.}, \& \bibinfo{author}{Puglisi, G.}
  (\bibinfo{year}{2017}).
\newblock \bibinfo{title}{Catastrophic thinning of dielectric elastomers}.
\newblock {\it \bibinfo{journal}{Phys. Rev. Lett.}\/},  {\it
  \bibinfo{volume}{118}\/}, \bibinfo{pages}{078001}.

\end{thebibliography}
 \end{document}